\documentclass[review,12pt]{elsarticle}

\usepackage{amssymb}
\usepackage{amsmath}
\usepackage{comment}
\usepackage{tabularx}
\usepackage{booktabs}
\usepackage{url}
\journal{Nuclear Physics B}

\begin{document}

\begin{frontmatter}

%% Title, authors and addresses

%% use the tnoteref command within \title for footnotes;
%% use the tnotetext command for theassociated footnote;
%% use the fnref command within \author or \affiliation for footnotes;
%% use the fntext command for theassociated footnote;
%% use the corref command within \author for corresponding author footnotes;
%% use the cortext command for theassociated footnote;
%% use the ead command for the email address,
%% and the form \ead[url] for the home page:
%% \title{Title\tnoteref{label1}}
%% \tnotetext[label1]{}
%% \author{Name\corref{cor1}\fnref{label2}}
%% \ead{email address}
%% \ead[url]{home page}
%% \fntext[label2]{}
%% \cortext[cor1]{}
%% \affiliation{organization={},
%%             addressline={},
%%             city={},
%%             postcode={},
%%             state={},
%%             country={}}
%% \fntext[label3]{}

\title{A Multimodal Human-Centered Framework for Assessing Pedestrian Well-Being in the Wild}
%A Naturalistic, Multimodal Framework for Understanding Pedestrian Well-Being

%% use optional labels to link authors explicitly to addresses:
%% \author[label1,label2]{}
%% \affiliation[label1]{organization={},
%%             addressline={},
%%             city={},
%%             postcode={},
%%             state={},
%%             country={}}
%%
%% \affiliation[label2]{organization={},
%%             addressline={},
%%             city={},
%%             postcode={},
%%             state={},
%%             country={}}

\author[label1]{Yasaman Hakiminejad} %% Author name
\author[label2]{Arash Tavakoli}
%% Author affiliation
\affiliation[label1,label2]{organization={Civil and Environmental Engineering, Villanova University},%Department and Organization
            addressline={800 E Lancaster Ave}, 
            city={Villanova},
            postcode={19085}, 
            state={PA},
            country={United States}}

%% Abstract
\begin{abstract}
Pedestrian well-being is a critical yet rarely measured component of sustainable urban mobility and livable city design. Existing approaches to evaluating pedestrian environments often rely on static, infrastructure-based indices or retrospective surveys, which overlook the dynamic, subjective, and psychophysiological dimensions of everyday walking experience. This paper introduces a multimodal, human-centered framework for assessing pedestrian well-being in the wild by integrating three complementary data streams: continuous physiological sensing, geospatial tracking, and momentary self-reports collected using the Experience Sampling Method. The framework conceptualizes pedestrian experience as a triangulation across environmental context, objective physiological response, and subjective perception, enabling a holistic and situated understanding of how urban environments influence well-being. The utility of our framework is then demonstrated through a naturalistic case study conducted in the Greater Philadelphia region, in which participants wore research-grade wearable sensors and carried GPS-enabled smartphones during their regular daily activities. Physiological indicators of autonomic nervous system activity, including heart rate variability and electrodermal activity, were synchronized with spatial trajectories and in situ self-reports of stress, affect, and perceived infrastructure conditions. Results illustrate substantial inter- and intra-individual variability in both subjective experience and physiological response, as well as context-dependent patterns associated with traffic exposure, pedestrian infrastructure quality, and environmental enclosure. The findings also suggest that commonly used walkability indices may not fully capture experiential dimensions of pedestrian well-being. By enabling real-world, multimodal measurement of pedestrian experience, the proposed framework offers a scalable and transferable approach for advancing human-centered urban analytics. 

\end{abstract}

%% Keywords

\end{frontmatter}

\section{Introduction}
\label{sec1}
%%% Introduction Paragraph Structure Guide

%%% 1. Context and Motivation
%%% Explain why pedestrian well-being is important in cities, 

As urban areas across the United States continue to expand, the importance of walkability has become increasingly emphasized in planning and design practices \cite{litman2003economic,abastante2020supporting}. Traditionally, infrastructure in the US has been designed with a primary focus on vehicular traffic, often neglecting the needs of pedestrians \cite{badami2009urban, budzynski2019effects, giacomini2017pedestrian}, cyclists \cite{guo2023psycho,guo2023rethinking}, and transit users \cite{hakiminejad2025public,hakiminejad2025understanding}. However, as the negative effects of car-centric urban development, such as traffic congestion and air pollution, become more evident, cities are increasingly recognizing the need to prioritize other modes of transportation such as pedestrian-friendly environments \cite{wang2025city, alwan2021measuring}. This shift in focus is not only aimed at improving transportation options but also at enhancing the overall quality of life for residents \cite{tavakoli2025psycho}. In simple terms, urban spaces that are designed with pedestrians in mind can lead to healthier, more active lifestyles, reduced environmental impact, and stronger community engagement \cite{calvert2019urban, distefano2023fostering}.

Emerging from this shift is a broader understanding of ``pedestrian well-being,'' which refers to the physical, emotional, and psychological aspects of walking in urban environments \cite{calvert2015urban, zacharias2001pedestrian, middleton2010sense}. Pedestrian well-being goes beyond safety and accessibility to encompass factors such as comfort, stress levels, social interaction, and the sensory experience of the urban landscape. It involves the ability to move through public spaces without undue anxiety, enjoy aesthetically pleasing surroundings, and feel physically comfortable while walking \cite{forsyth2010promoting, gehl2013cities}. Well-being is also shaped by social and emotional connections to the environment—how people feel in their neighborhood, how safe they feel while walking, and how the design of public spaces encourages or discourages social interaction \cite{park2020pedestrian,lee2017relation}. \textbf{Overall, pedestrian well-being represents a holistic construct that integrates the physical and emotional aspects of the walking experience, emphasizing the creation of environments that support not only the mobility of individuals but also their overall quality of life.}

Pedestrian well-being is influenced by a range of environmental factors, including safety, accessibility, comfort, and aesthetics. These dimensions are shaped by urban design elements such as sidewalk quality, crossing infrastructure, greenery, and seating, which collectively influence perceptions of safety and comfort during walking \cite{forsyth2010promoting, gehl2013cities}. Despite growing recognition of these factors, existing tools used to evaluate pedestrian environments, such as walkability indices, often rely on static, aggregated data that may not reflect real-time or subjective experiences \cite{leslie2007walkability, frank2010development}. Common indices such as Walk Score and Pedestrian Level of Service ratings tend to focus on functional criteria, such as access to amenities, street connectivity, and traffic dynamics \cite{carr2010walk, duncan2011validation, lo2009walkability}. While useful, these measures can obscure the nuanced, context-dependent aspects of pedestrian well-being and may overlook critical subjective factors, including perceived safety, emotional responses, comfort, and aesthetic quality \cite{alfonzo2005walk}. In addition, these tools typically apply generalized scoring systems that overlook local context, seasonal variability, and user diversity. For example, a neighborhood might receive a high rating due to its proximity to commercial destinations, yet still degrade pedestrian well-being due to deteriorated sidewalks, excessive noise, or lack of shade and seating \cite{forsyth2015walkable}. \textbf{These limitations underscore the need for more responsive and dynamic approaches to capturing and understanding the pedestrian experience.}

This paper proposes a multimodal framework for assessing pedestrian well-being that conceptualizes the problem as a triangulation across three data streams: physiological sensing, geospatial tracking, and subjective self-reports. By integrating these modalities, objective biosignals (e.g., electrodermal activity, heart rate), spatial context (e.g., GPS trajectories), and self-reported emotional or cognitive states, the framework offers a more comprehensive understanding of pedestrians’ lived experiences within urban environments. This approach enables the identification of latent stress patterns and environmental triggers that are often missed in single-modality studies. Each component is designed for real-world, longitudinal deployment, making it possible to capture situated and authentic responses to urban infrastructure over time.

To evaluate the feasibility and applicability of the proposed multimodal framework, a real-world case study was conducted in the Greater Philadelphia region. Participants were recruited and equipped with wearable physiological sensors, GPS-enabled smartphones, and a mobile survey interface, which they used over multiple days during their regular daily activities. This naturalistic deployment enabled the continuous collection of (1) electrodermal activity and heart rate variability to quantify physiological states, (2) geospatial trajectories to contextualize location-based patterns, and (3) brief self-reports to capture in situ measures of subjective well-being, including perceived stress, arousal, valence, and mental load. Unlike structured route studies, this approach captures authentic, unconstrained pedestrian behavior across diverse urban settings. This deployment serves as an initial validation of the framework’s capacity to detect situated, multimodal indicators of pedestrian well-being in real-world contexts. The multimodal approach offers a new lens for advancing human-centered urban design by directly linking the physiological, spatial, and psychological dimensions of pedestrian experience.

\section{Background Information}
Advances in wearable technologies, mobile-based data collection, and geolocation tracking have enabled new approaches for evaluating transportation experiences across multiple modes, including walking, cycling, and public transit \cite{wong2021wearable,marquart2022experiences} beyond survey methods \cite{naseralavi2025machine}. These tools enable researchers to move beyond static, infrastructure-focused assessments and instead capture the real-time, lived experiences of transportation users \cite{vieira2011smart,chow2016utilizing}.

Wearable physiological sensors have proven particularly effective for assessing emotional and physical responses to the built environment. Devices that measure heart rate variability (HRV), electrodermal activity (EDA), electroencephalography (EEG), and skin temperature provide insights into stress, arousal, and physical exertion as individuals navigate different urban contexts \cite{kyriakou2019detecting, pykett2020urban, saitis2018multimodal}. Recent studies have demonstrated their practical application. For instance, Resch et al. integrated wearable sensors with geospatial data to map ``urban stress hotspots'' in Salzburg, revealing how specific features such as noise, traffic density, and narrow sidewalks triggered measurable physiological stress in pedestrians \cite{resch2020interdisciplinary}. Similarly, Gidlow et al. used heart rate monitors and cortisol measurements to demonstrate that urban green spaces were associated with reduced stress compared to built-up environments, underscoring the health benefits of restorative settings \cite{gidlow2016put}. Mavros et al. employed mobile EEG and EDA to capture pedestrians’ emotional responses in natural urban environments, proposing an integrated framework linking affective states to spatial behavior \cite{mavros2019measuring}. Hogertz et al. also used EDA to monitor pedestrians walking predefined routes in Lisbon, showing that specific environmental features such as traffic, noise, and narrow sidewalks elicited measurable changes in arousal \cite{hogertz2010emotions}. Beyond walking, researchers have also applied these devices across other transportation modes. For example, EDA has been used to assess stress and emotional responses in diverse real-world mobility contexts, including monitoring drivers’ physiological stress in complex urban environments \cite{venkatachalapathy2022naturalistic}, evaluating emotional states during routine driving \cite{milardo2021understanding}, and detecting stress in motorcyclists navigating dense traffic \cite{corcoba2017prediction}.

A primary limitation of using wearable devices in isolation is that, although they provide rich physiological insights, they lack contextual and subjective interpretation. This limitation makes it difficult to determine whether observed physiological changes reflect stress, exertion, or other affective responses without complementary self-report data. Mobile experience sampling methods (ESM), typically delivered via smartphone applications, complement these physiological measures by capturing subjective perceptions during or immediately after transport activities. Unlike retrospective surveys, mobile-based ESM allows for the real-time reporting of mood, safety, satisfaction, and environmental comfort. Numerous studies, both within and beyond walking research, have employed ESM to capture real-time subjective experiences. For instance, Bakolis et al., in the ``Urban Mind'' project, used this approach to assess how exposure to natural elements such as trees and water influenced the mental well-being of London pedestrians and transit users, finding strong associations between greenery and improved mood states \cite{bakolis2018urban}. 

Global Positioning System (GPS) tracking further enhances understanding of mobility experiences by precisely mapping movement patterns. GPS data allow researchers to analyze route choices, dwell times, and avoidance behaviors across different urban contexts. Shen and Stopher reviewed the application of GPS in transportation studies, highlighting its advantage over self-reported travel logs in capturing accurate spatiotemporal data, especially for complex multimodal trips \cite{shen2014review}. In a related example, Hood et al. used GPS data to examine route choice among cyclists in Philadelphia and found that many riders were willing to take longer routes in exchange for dedicated bike lanes, suggesting that perceptions of safety and comfort significantly influenced behavior \cite{hood2011gps}.

When these technologies are integrated in the same framework, they provide a multidimensional view of the transportation experience. For instance, within driving research, Tavakoli et al. introduced the ``HARMONY framework,'' a naturalistic driving study that collected synchronized multimodal data, including heart rate, GPS, and ambient stimuli to analyze driver states and their contextual triggers across time and space \cite{tavakoli2021harmony}. Such multimodal methodologies enable researchers and urban planners to move beyond conventional infrastructure audits toward designing transportation systems that incorporate both objective conditions and subjective human experiences.

Within pedestrian research, however, the integration of multimodal sensing with contextual and perceptual data remains relatively limited. Although a few studies have begun to bridge this gap, they remain limited in scope or methodologically isolated. For example, Dörrzapf et al. introduced the \textit{``Walk \& Feel''} framework, which integrates biosensor data with environmental and subjective inputs to assess pedestrian perception and emotional responses to different streetscapes \cite{dorrzapf2019walk}. The authors proposed a methodology that combined physiological measurements, such as skin conductance, heart rate variability, and skin temperature, with self-reported and spatial data to examine emotional reactions to the built environment. Similarly, Kim et al. demonstrated that large-scale biosignal data from wearables, including electrodermal activity and heart rate, can be used to predict spatially anchored environmental distress, highlighting the value of real-world, multimodal data integration in understanding pedestrian discomfort and safety perception \cite{kim2023location}. They combined geocoded physiological signals with self-reported environmental feedback to model collective distress among pedestrians in urban contexts. Using spatial analytics and machine learning, the model achieved 80\% prediction accuracy, revealing strong associations between biosignals and negative urban stimuli such as neglected infrastructure, noise, and crowding. Saitis and Kalimeri (2018) proposed a multimodal biosensing framework combining EEG, EDA, and HRV to assess stress and cognitive load in visually impaired pedestrians navigating both indoor and outdoor environments \cite{saitis2018multimodal}.  Johnson, Kanjo, and Woodward (2023) introduced the \textit{``DigitalExposome''} framework, a multimodal sensing approach designed to quantify how urban environmental factors influence human well-being. Their study integrated real-time measurements of air pollutants (e.g., PM$_1$, PM$_{2.5}$, PM$_{10}$, NH$_3$), noise, and crowd density with physiological biosignals such as EDA, heart rate, HRV, and body temperature, along with self-reported emotional states in a predefined urban path \cite{johnson2023digitalexposome}.

%%%%need to have a paragrph mentioning the gaps of those studies and how our study overcomes it. 
Although previous studies have shed light on how urban environments shape pedestrian stress, affect, and behavior, existing work remains fragmented across sensing modalities and rarely integrates multiple data streams in a unified framework. While recent work in multimodal simulation has begun to address similar challenges through the development of human-centered, immersive platforms \cite{azimi2025simulation}, much of the literature relies on single-sensor physiological measurements (e.g., HRV or EDA alone) or on short, predefined walking routes, limiting the ecological validity needed to understand real-world pedestrian experience. Even studies that adopt multimodal sensing, such as those combining EEG and EDA, or HRV, EDA, and environmental exposures, often remain methodologically siloed, analyzing each signal independently rather than aligning physiological data, subjective perception, and spatial context over time. Moreover, most frameworks lack continuous, naturalistic GPS tracking and seldom incorporate in-situ Experience Sampling to contextualize physiological changes with subjective reports of stress, affect, or environmental discomfort. As a result, we still have limited understanding of how moment-to-moment autonomic responses correspond to the specific physical, perceptual, and contextual features of everyday walking environments.

\begin{figure}[h!]
\centering
\includegraphics[width=0.9\textwidth]{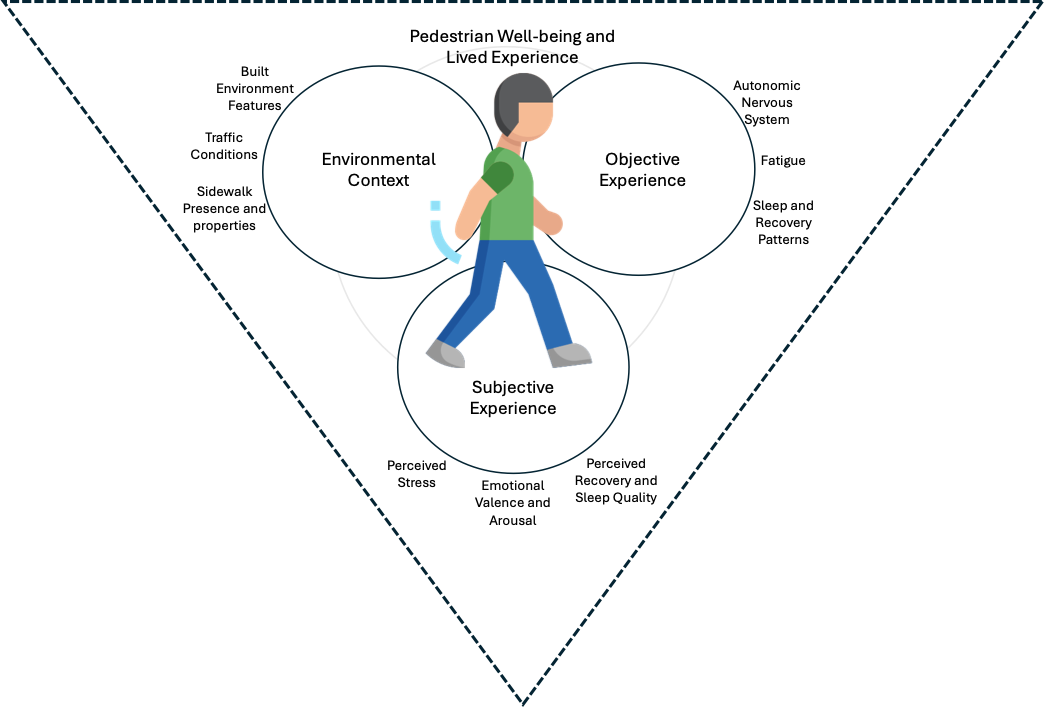}
\caption{Triangulated conceptual model of pedestrian well-being. The model integrates three interrelated dimensions—Environmental Context, Objective Experience, and Subjective Experience—to holistically capture the lived experience of pedestrians in urban environments.}
\label{fig:triangle}
\end{figure}

This study addresses these gaps by introducing a unified, human-centered multimodal framework that synchronizes three complementary data sources: continuous physiological signals (HRV, EDA, skin temperature, accelerometry), GPS-based environmental context, and momentary self-reports via ESM. The framework adopts a triangulated perspective that integrates three key dimensions: Environmental Context, Objective Experience, and Subjective Experience (Figure~\ref{fig:triangle}). Environmental context refers to the physical and spatial features of the built environment, such as infrastructure, traffic conditions, and green space. Objective experience encompasses physiological and behavioral indicators of well-being, including stress-related biomarkers and physical activity. Subjective experience captures momentary self-reports of emotional and cognitive states. These three aspects collectively shape pedestrian well-being because each provides a fundamentally different lens on the walking experience: environmental context identifies where people move, objective experience captures how their bodies respond, and subjective experience reveals how they interpret and feel about those moments. Considering these dimensions together enables the framework to move beyond isolated measurements and to construct a holistic, situated understanding of pedestrian experience that no single modality could provide alone.

Unlike prior work, the proposed approach is deployed in real-world settings, capturing unstructured and spontaneous walking behavior across diverse urban environments rather than within constrained laboratory or route-based protocols. By integrating individualized physiological baselines, fine-grained spatial tracking, and real-time subjective report, the framework enables a richer, more context-sensitive understanding of pedestrian well-being. This triangulated approach provides the methodological foundation for identifying how everyday environmental conditions, including traffic exposure, sidewalk quality, enclosure, and sensory load, may influence both objective autonomic responses and subjective emotional states during real-world walking.

%%%%%%%%%%%%%%%%%%%all the previously commented material is below%%%%%%%%%%
%Another study used qualitative interviews and walking diaries to explore pedestrians’ cognitive, emotional, and sensory experiences \cite{calvert2015urban} these approaches are typically conducted independently. Comprehensive methodologies that synthesize subjective feedback, biometric signals, and geospatial analytics in a unified framework remain underdeveloped in practice.

% This gap underscores a pressing need to develop robust, scalable frameworks that integrate these three modalities (physiological, spatial, and subjective) to more meaningfully evaluate and design urban environments that promote psychological comfort, health, and overall pedestrian well-being.

%Another study performed by Venkatachalapathy etl al. measured cyclists’ physiological stress in response to car passing, road markings, and parked vehicles using wearable EDA devices during naturalistic biking trips in Iowa \cite{venkatachalapathy2022naturalistic}.

\section{Framework Components}
Building on this conceptual foundation (Figure ~\ref{fig:triangle}), we propose a multimodal framework for assessing the interaction between pedestrian well-being and walking infrastructure (Figure~\ref{fig:framework}). This framework collects data across multiple dimensions of well-being, including stress, emotional state, sleep quality, and physical activity. To triangulate pedestrian experience, the study integrates data from three distinct modules: (1) physiological sensing via smartwatches, (2) self-reported momentary assessments through the Experience Sampling Method (ESM), and (3) geospatial tracking through GPS. Together, these data sources provide a comprehensive view of how pedestrians experience and respond to their environments in real time. The components of this framework are shown in Figure \ref{fig:framework} and are detailed as follows: 

\begin{figure}[h!]
\centering
\includegraphics[width=0.9\textwidth]{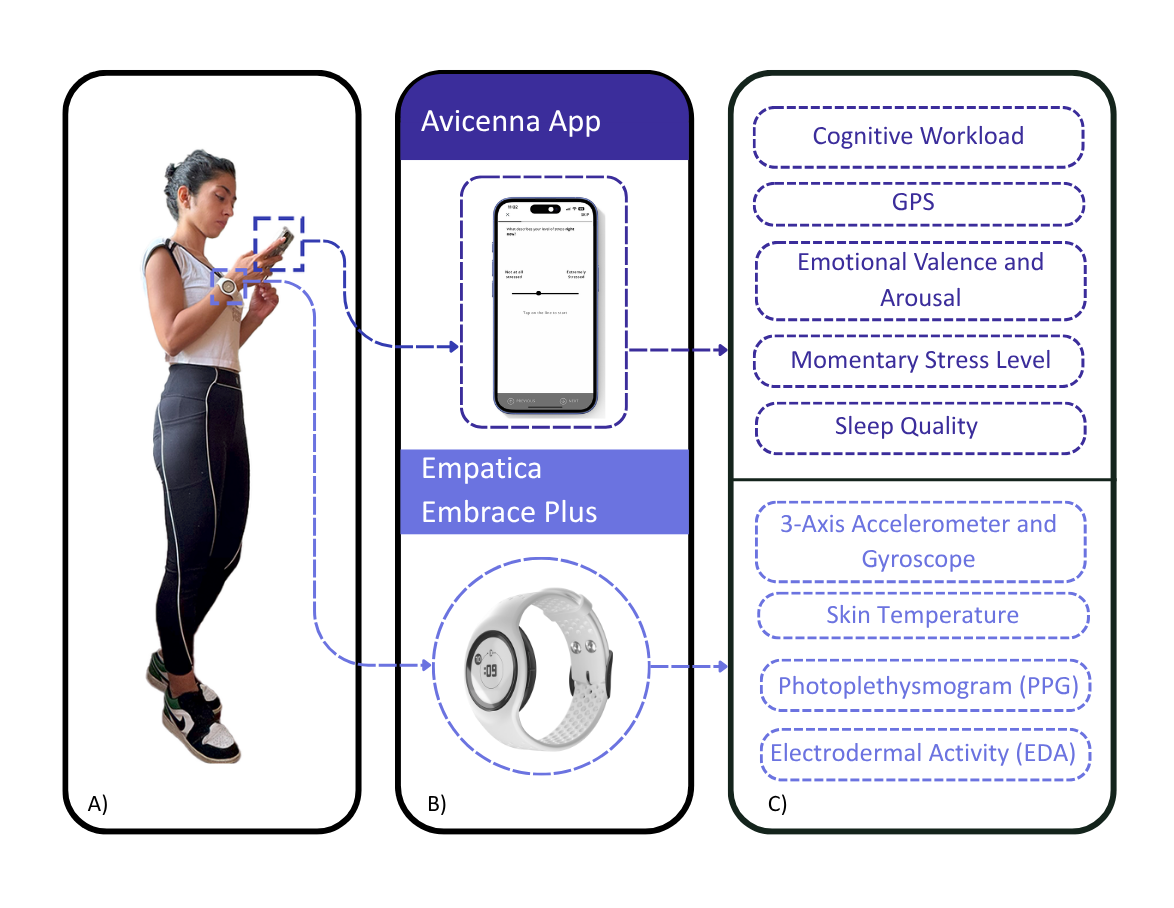}
\caption{Overview of the data collection framework used in the study.
(A) A participant engages with the study app while wearing the Empatica EmbracePlus smartwatch.
(B) The Avicenna App administers self-report measures using the Experience Sampling Method (ESM), while the Empatica device continuously collects physiological signals.
(C) Data streams include subjective responses (e.g., momentary stress, emotional valence/arousal, sleep quality), GPS-based location tracking, and sensor-based physiological signals (e.g., accelerometry, skin temperature, photoplethysmography (PPG), and electrodermal activity (EDA)). Together, these modalities enable a high-resolution, multimodal assessment of real-world well-being and environmental context during urban walking.}
\label{fig:framework}
\end{figure}

\subsection{Physiological Signal Acquisition}

This module captures participants’ objective physiological and behavioral metrics, offering insight into their moment-to-moment experiences in the walking environment. Specifically, it allows for the assessment of autonomic nervous system activity, such as sympathetic arousal and parasympathetic recovery, using indicators such as heart rate and heart rate variability, which are closely linked to stress and emotional regulation \cite{tavakoli2021harmony,kim2018stress}. In addition to physiological responses, this module tracks participants’ movements and physical activity levels, enabling analysis of walking behavior and mobility patterns. The module is designed to be hardware-agnostic and can operate independently of any specific device, as long as the device supports continuous acquisition of relevant biosignals such as heart rate, accelerometry, or electrodermal activity. In our deployment, we used the Empatica EmbracePlus as a representative example of a research-grade wearable sensor due to its capability to collect key signals, including blood volume pulse (BVP), electrodermal activity, skin temperature, and 3-axis accelerometry \cite{embraceplus}.

\subsection{Geospatial Tracking Module}
The geospatial tracking module is designed to provide continuous contextual information about a participant’s movement and environmental exposure. This module performs  periodic sampling of GPS coordinates comprising latitude, longitude, timestamp, and speed. In this deployment, we implemented the module using the Avicenna Research mobile application (formerly known as Ethica), a research-grade tool for passive behavioral data collection \cite{avicennaplatform}. 

\subsection{Subjective Experience Module}

The subjective experience module is designed to capture momentary self-reports of emotional and behavioral states in real-world contexts using the Experience Sampling Method, to construct the subjective experience portion of the framework. ESM is a widely validated technique for assessing mood, cognition, and contextual experience in situ~\cite{larson2014experience}. This module is flexible and reproducible using any digital platform capable of delivering time-triggered surveys and logging responses with temporal precision.

To capture a wide variety of subjective experiences, this module included the following standardized questionnaires:
\begin{itemize}
\item \textbf{Perceived stress level}: Participants rated their current level of stress on a scale from 0 (no stress at all) to 10 (worst stress possible) in response to the prompt: ``What describes your level of stress right now?'' \cite{karvounides2016three}

\item \textbf{Emotional state}: Arousal was measured by asking, ``How activated do you feel right now? Activated means charged, alert, or energized physically and mentally,” with responses ranging from ``Not at all activated'' to ``Extremely activated.'' Valence was assessed using the question, ``How positive, neutral, or negative do you feel right now?'' rated on a scale from ``Extremely negative” to “Extremely positive.'' \cite{mauss2009measures}

\item \textbf{Sleep quality}: Each morning, participants retrospectively rated their sleep using the prompt: ``How would you rate the quality of your sleep overall last night?'' on a 5-point scale from ``Terrible” to “Excellent.'' \cite{krystal2008measuring}

\item \textbf{Walking behavior}: At the end of each day, participants reported whether they had walked and for how long.

\item \textbf{Observations of pedestrian infrastructure}: Participants were asked, ``Did you notice any issues with pedestrian infrastructure during your walk? (please elaborate).'' providing open-ended feedback on perceived environmental barriers or discomfort.
\end{itemize}

\section{Case Study: Implementation in the Greater Philadelphia Region}

To evaluate the applicability and potential insights of the proposed framework, this case study implements the multimodal approach in the Greater Philadelphia region to examine how urban walking environments influence pedestrian well-being. Rather than focusing solely on the feasibility of data collection, the study aimed to investigate how integrating physiological signals, geospatial information, and subjective self-reports could reveal dynamic interactions between pedestrians and their surrounding environments. Applying the framework in a real-world setting allowed the study to capture the lived experiences of pedestrians across diverse urban contexts and to assess how moment-to-moment well-being fluctuates in relation to environmental and contextual factors.

% \subsection{Study Overview}

% To this end, we recruited 34 participants from the Greater Philadelphia region to ensure ecological validity while maintaining logistical feasibility for deployment and monitoring. Figure~\ref{fig:gps_map} displays the geographic distribution of area of study captured during walking episodes. The data span urban, suburban, and peri-urban areas of the Greater Philadelphia region, providing diverse environmental contexts for analysis.
% Data collection took place over a two-week period. During this time, each participant contributed continuous physiological and GPS data and completed four daily surveys assessing stress, mental load, arousal, and valence, along with a daily report on sleep quality.

\subsection{Participant Recruitment and Eligibility Criteria}
Participant recruitment was based on three primary inclusion criteria. First, participants were required to regularly engage in walking as a mode of transportation (e.g., for commuting, attending classes, running errands, or exercising) to ensure adequate exposure to varied pedestrian infrastructure. Second, participants needed to own a smartphone compatible with the study’s mobile applications for GPS tracking, physiological data collection, and survey delivery. All applications were designed to ensure privacy by excluding personally identifiable information. Third, participants were instructed to wear a biosignal monitoring device (e.g., Empatica EmbracePlus) during waking hours to enable continuous tracking of physiological indicators related to well-being and environmental stress.

% Participant recruitment was a critical component of the study design; three primary inclusion criteria were applied to determine eligibility. First, participants were required to incorporate walking as a primary or routine mode of transportation. This included individuals who walked as part of their daily commute, to classes, or for other regular activities such as errands or exercise. This criterion was essential to ensure adequate exposure to a variety of pedestrian infrastructure. Second, participants needed to own a personal smartphone compatible with the study’s mobile applications. They were required to download and run apps that enabled GPS tracking, physiological data recording, and survey delivery. Importantly, the selected apps did not collect any personally identifiable information, ensuring participant privacy and data security. Third, participants were asked to wear a biosignal monitoring device (e.g., Empatica EmbracePlus) for as many waking hours as possible throughout the study period. This continuous physiological monitoring was necessary to capture moment-to-moment variations in autonomic nervous system activity, which served as key indicators of well-being and environmental stress. 

A total of 34 participants were recruited from a university community, comprising primarily undergraduate and graduate students. All participants provided written informed consent prior to participation, in accordance with the study protocol approved by the Institutional Review Board (IRB approval number: IRB-FY2024-218), and were informed of the study duration, data collection procedures, and potential risks before enrollment.

Participants had a mean age of 23.6 years (SD = 6.5; range = 18–50). The sample included 20 participants (58.8\%) identifying as male and 14 (41.2\%) identifying as female. A summary of participant characteristics is provided in Table~\ref{demographics}.

\begin{table}[ht!]
\centering
\caption{Demographics of participants (N=34)}
\label{demographics}
\resizebox{0.85\textwidth}{!}{%
\begin{tabularx}{\textwidth}{@{}Xccc@{}}
\toprule
\textbf{Variable} & \textbf{Attribute} & \textbf{Number} & \textbf{Percent (\%)} \\
\midrule
\textbf{Gender}  & Male & 20 & 58.82 \\
& Female & 14 & 41.18 \\
\midrule
\textbf{Ethnicity} 
& White & 17 & 50.00 \\
& Asian & 8 & 23.53 \\
& Black or African American & 4 & 11.76 \\
& Other (please specify) & 4 & 11.76 \\
& Prefer not to say & 1 & 2.94 \\
\midrule
\textbf{Education} 
& Some college & 14 & 41.18 \\
& Advanced degree (M.A., M.S., Ph.D., M.D., J.D.) & 9 & 26.47 \\
& Bachelor's degree (B.A., B.S.) & 5 & 14.71 \\
& High school degree & 4 & 11.76 \\
& Associate's degree (A.A., A.S.) & 2 & 5.88 \\
\midrule
\textbf{Age} & Mean (Std. Dev) & 23.62 (6.53) & -- \\
 & Range & 18--50 & -- \\
\bottomrule
\end{tabularx}%
}
\end{table}

\subsection{Instrumentation and Devices}

Each participant was provided with an Empatica EmbracePlus smartwatch, a research-grade wearable equipped with multiple physiological sensors: a photoplethysmography sensor, an electrodermal activity sensor, a skin temperature sensor, and a 3-axis accelerometer. The PPG sensor captures blood volume pulse, from which heart rate and heart rate variability (HRV) are derived, key indicators of autonomic nervous system activity. The EDA sensor measures skin conductance, providing indicators of sympathetic arousal and emotional reactivity. The skin temperature sensor records peripheral temperature fluctuations, which may reflect changes in stress or metabolic state. The 3-axis accelerometer monitors physical movement and activity intensity. Together, these sensors enable continuous, multimodal monitoring of participants’ physiological states and behaviors. Data were continuously synchronized with participants’ smartphones via the Empatica mobile application, ensuring secure and passive transmission throughout the study period. Physiological data were collected continuously through the Empatica EmbracePlus smartwatch and transmitted via the Carelab app installed on participants’ smartphones. The application recorded and securely uploaded raw sensor data, including heart rate and electrodermal activity, to a cloud server identified by a unique participant ID.

Through using the Avicenna app, participants received four brief surveys per day, delivered at random times within fixed time windows (morning, afternoon, evening, and night). Each survey was structured to assess both momentary and retrospective aspects of well-being and behavior, including perceived stress, valence, arousal, and contextual factors. Using the same application, GPS data were collected to provide a geospatial record of participant mobility. To balance temporal resolution and battery life, GPS data were recorded at five-minute intervals. GPS data were stored locally and securely uploaded to a cloud-based server once an internet connection was available. Each GPS sample included timestamped spatial coordinates and associated metadata such as speed.

All data were anonymized at the point of collection using non-identifiable participant IDs. No personally identifiable information was accessible to third-party platforms (Empatica and Avicenna), ensuring participant privacy throughout the data pipeline.
Participants were compensated \$50 upon completion of the two-week study period.

\subsection{Data Preprocessing and Synchronization}

To integrate multimodal data streams and enable meaningful analysis, all physiological, geospatial, and self-reported data underwent structured preprocessing and temporal alignment. Given the asynchronous nature of these data sources, physiological signals sampled continuously, GPS coordinates recorded intermittently, and surveys completed at discrete time intervals, synchronization was essential to ensure temporal coherence across modalities. This section details the preprocessing steps applied to each data modality, including signal cleaning and feature extraction from physiological recordings, filtering and alignment of GPS traces, and processing and categorization of ESM responses.

\subsubsection{Heart Rate Variability}

%For brevity, in the following case study, we utilize only the PPG sensor from the Empatica EmbracePlus smartwatch to extract physiological indicators.

The analysis first focused on heart rate variability metrics and feature extraction, followed by the examination of electrodermal activity features.
HRV is a widely used, non-invasive marker of autonomic nervous system (ANS) function. Specifically, the analysis employed the Root Mean Square of Successive Differences (RMSSD), a standard time-domain HRV metric that primarily reflects parasympathetic (vagal) activity \cite{laborde2017heart}. RMSSD is calculated as the square root of the mean of the squared differences between successive inter-beat intervals (IBIs), which are typically derived from R–R intervals in ECG or systolic peak intervals in PPG. RMSSD was computed using the following equation:

\begin{equation}
\mathrm{RMSSD} = \sqrt{\frac{1}{N-1} \sum_{i=1}^{N-1} (IBI_{i+1} - IBI_i)^2}
\end{equation}

where $IBI_i$ represents the $i^{th}$ inter-beat interval, and $N$ is the total number of intervals in the analysis window. The IBI series was provided by Empatica as part of the device output, although it can also be derived directly from raw PPG signals using standard signal processing techniques.

% Physiological data were collected using the Empatica EmbracePlus wearable device, which records blood volume pulse (BVP) at 64 Hz. Raw BVP signals were exported from Empatica’s cloud platform and processed offline. For each participant, systolic peaks were identified using either (1) the device-provided systolic peak timestamps when available, or (2) a custom peak detection algorithm applied to the raw BVP waveform in MATLAB.

Following established procedures in prior research \cite{bernardes2022reliable,pham2021heart,uryga2025impact}, RMSSD was computed using a 5-minute rolling window with a 1-second step size, producing a high-resolution time series of HRV estimates across the entire study period. The computation was implemented using custom Python scripts to ensure consistent windowing and temporal alignment across participants and study days. The resulting RMSSD time series served as the primary physiological indicator of momentary autonomic regulation throughout the study. To identify periods of elevated physiological stress, RMSSD values falling below the 5th percentile of each participant’s daily distribution during walking periods were classified as ``arousal episodes,'' following percentile-based thresholding approaches commonly used in physiological signal processing contexts \cite{jacobson2001auto, rahimzadeh2024significance}.

\subsubsection{Electrodermal Activity Analysis}

Electrodermal activity constituted the second physiological data source and was continuously recorded via the Empatica EmbracePlus smartwatch, which sampled skin conductance from the ventral wrist at 4 Hz. Raw data were exported from Empatica’s cloud platform and analyzed offline using Ledalab (v3.4.9), a validated MATLAB-based toolbox for psychophysiological data analysis \cite{benedek2010continuous, ledalab_doc}.

Following standard procedures used in prior EDA research \cite{boucsein2012electrodermal, dawson2007electrodermal, kreibig2010autonomic}, continuous decomposition analysis (CDA) was performed for each participant and session using Ledalab’s implementation of a bi-exponential impulse response function (IRF), also known as the Bateman function:

\begin{equation}
\text{IRF}(t) = \frac{1}{\tau_2 - \tau_1} \left( e^{-t/\tau_2} - e^{-t/\tau_1} \right)
\end{equation}

where $\tau_1$ and $\tau_2$ represent the rise and recovery time constants of skin conductance responses (SCRs), respectively. Default parameter values were $\tau_1 = 0.7$~s and $\tau_2 = 2.0$~s. Level 4 parameter optimization in Ledalab was enabled to refine IRF parameters and maximize the compactness and non-negativity of the estimated phasic driver \cite{benedek2010continuous, alexander2005separating}.

After decomposition, the tonic and phasic components were exported and further analyzed using custom Python scripts. The EDA signal was segmented into overlapping 5-minute windows with a 1-second step size, producing a high-resolution time series of autonomic features. Within each window, both tonic and phasic metrics commonly used in psychophysiological research were extracted \cite{boucsein2012electrodermal, dawson2007electrodermal, kreibig2010autonomic}.

The following features were computed from the phasic component:

\begin{itemize}
  \item \textbf{Standard Deviation of Phasic Driver:} Quantifies the variability of short-term sympathetic activation within each analysis window.
  \item \textbf{Integrated Skin Conductance Response (ISCR):} Total phasic activation calculated as the integral over the window:
  \begin{equation}
  \text{ISCR} = \int_{t_0}^{t_1} \text{Driver}_{\text{phasic}}(t) \, dt
  \end{equation}
  This measure reflects cumulative sympathetic output and is robust to overlapping responses \cite{benedek2010continuous}.
  \item \textbf{Number of Significant SCRs:} Count of discrete SCR events with amplitudes $\geq 0.02~\mu$S, following standard detection thresholds \cite{dawson2007electrodermal}.
  \item \textbf{SCR Frequency:} Number of significant SCRs per minute, reflecting sympathetic reactivity.
  \item \textbf{Maximum SCR Amplitude:} The largest SCR magnitude observed within each analysis window.
  \item \textbf{Total SCR Amplitude Sum:} Sum of amplitudes of all significant SCRs, representing overall sympathetic activation.
\end{itemize}

In addition to phasic features, a tonic arousal metric—mean Skin Conductance Level (SCL), was also extracted, defined as the average of the tonic component within each window, reflecting slow-varying baseline arousal.

All decomposition parameters, window settings, and detection thresholds were kept constant across participants and sessions to ensure analytic consistency and reproducibility. This processing pipeline yielded temporally resolved, continuous measures of autonomic arousal suitable for subsequent statistical and machine learning analyses.
To identify periods of elevated sympathetic activation, we focused on the SCR frequency feature, which provides a robust index of short-term sympathetic reactivity. Following individualized thresholding approaches commonly used in signal processing context \cite{genovese2002thresholding}, we classified time windows in which SCR frequency exceeded the 95th percentile of each participant’s daily distribution as \textit{sympathetic arousal episodes}. This percentile-based method allowed the detection of moments characterized by relatively high sympathetic activity compared to each participant’s typical daily range. Thresholding was performed separately for each day to account for inter-day variability in baseline skin conductance and to maintain within-subject sensitivity to context-dependent fluctuations.

\subsubsection{GPS Data Filtering and Alignment}

Raw GPS data collected via the Avicenna mobile application exhibited irregular sampling intervals, with multiple entries recorded milliseconds apart despite a nominal sampling frequency of once every five minutes. To address this, the GPS logs were filtered by retaining only the first unique timestamp within each cluster of near-duplicate entries. This preprocessing step removed redundant points while preserving the temporal structure necessary for subsequent analyses.

Following deduplication, GPS data were filtered by speed to isolate walking behavior. Only data points corresponding to instantaneous speeds between 0.5 and 2.0 meters per second were retained, representing a slightly expanded range of typical walking speeds to account for minor fluctuations at the boundaries of walking episodes \cite{knoblauch1996field}. Consecutive data points within this range were grouped into walking segments lasting at least five minutes. For each segment, corresponding RMSSD values, EDA phasic features, and GPS coordinates were extracted and temporally aligned. This procedure ensured that geospatial data reflected ambulatory activity and were synchronized with physiological measures across participants and study days.

\begin{figure}[h]
\centering
\includegraphics[width=0.95\textwidth]{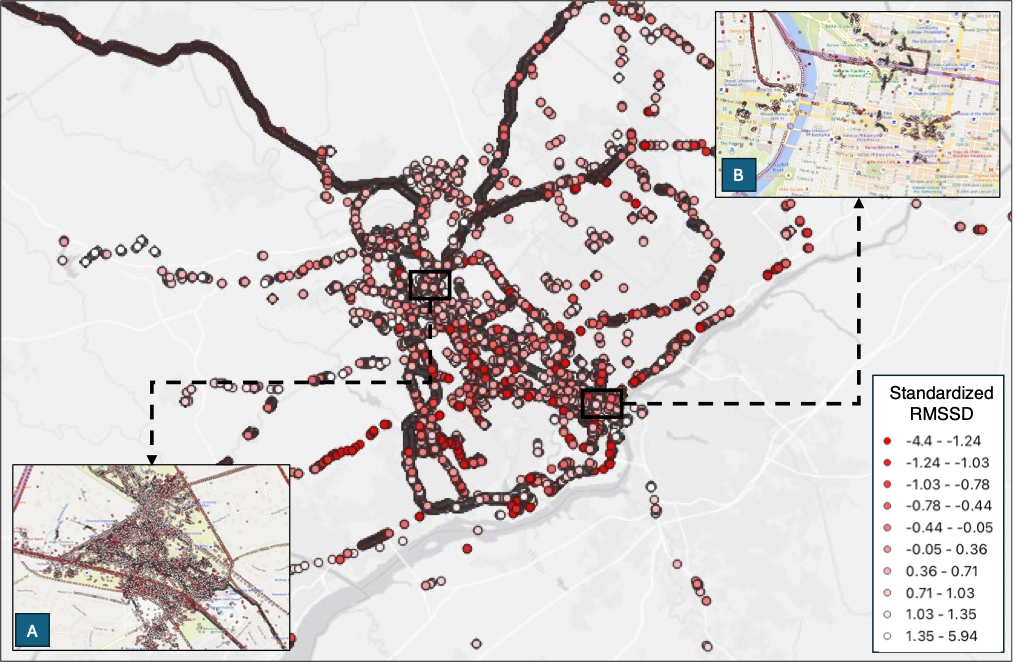}
\caption{Map of GPS traces from all participants, covering central Philadelphia and surrounding suburbs. Each GPS point is colored based on the participant’s standardized RMSSD value, with lower RMSSD (indicating less calm states) shown in darker red.}
\label{fig:gps_map}
\end{figure}

Figure~\ref{fig:gps_map} illustrates the integration of GPS data with RMSSD values and EDA phasic features, visualizing spatial variability in participants’ physiological states across walking routes in the Greater Philadelphia region. Each GPS point is color-coded according to the participant’s standardized RMSSD value, highlighting fluctuations in autonomic activity across space. The figure reveals substantial spatial variation, suggesting dynamic changes in participants’ physiological states as they traverse different urban environments. This spatialized representation provides an initial demonstration of how the multimodal framework captures nuanced patterns of pedestrian experience in real-world settings.

In addition to GPS-based physiological mapping, an environmental covariate representing local walkability was incorporated using the U.S. Environmental Protection Agency’s (EPA) Walkability Index \cite{epa2021walkability}. This index quantifies the relative ease of walking within an area by integrating multiple built environment characteristics following EPA User Guidelines \cite{epa2021walkability}. The Walkability Score ($W$) was calculated as a weighted sum of four standardized variables:

\begin{equation}
W = \left( \frac{w}{3} \right) + \left( \frac{x}{3} \right) + \left( \frac{y}{6} \right) + \left( \frac{z}{6} \right)
\end{equation}

where $w$ represents the block group’s ranked score for \textit{Intersection Density} (number of intersections per square mile), $x$ denotes the ranked score for \textit{Proximity to Transit Stops} (distance to the nearest bus, train, or rail stop), $y$ corresponds to the ranked score for \textit{Employment Mix} (entropy-based measure of job diversity), and $z$ represents the ranked score for \textit{Employment and Household Mix} (a composite of employment and residential density). Each variable was normalized to a national ranking scale prior to weighting, ensuring comparability across spatial contexts.

Walkability scores were retrieved from the EPA’s Smart Location Database for the Philadelphia metropolitan area and spatially joined to the corresponding GPS coordinates in the dataset.

\subsubsection{Survey Response Handling and Merging}

Four experience sampling surveys were administered daily, corresponding to different times of day. Survey data were first merged by participant ID to create a unified dataset per participant. To ensure data quality, individual survey entries containing missing or empty responses were excluded on a per-survey-type basis; however, no participants were removed entirely from the dataset. Response rates were calculated for each participant and survey type across the 14-day study period to quantify temporal coverage and participant engagement.

For the end-of-day survey, a keyword-based analysis was performed on responses to the pedestrian infrastructure question. Neutral or irrelevant responses (e.g., ``none,'' ``no,'' ``I didn’t see any issue'') were first removed. Remaining entries were categorized into predefined groups using keyword-based filtering. The categories were as follow: ``Sidewalk Issues,'' ``Crosswalk Issues,'' ``Uneven Surfaces,'' ``Trash and Debris,'' and ``Others''. All categorized responses were manually reviewed to ensure accuracy and to capture nuanced or misclassified entries not fully addressed by automated keyword rules.

% \subsection{Ethical Considerations}
% \subsubsection{IRB Approval and Oversight}
% \subsubsection{Data Security and Participant Privacy}
% -if any

\subsection{Results}
In the following subsections, we present findings derived from each data source within the proposed framework. The objective of this section is not to draw population-level conclusions, but rather to demonstrate the feasibility of integrating multimodal physiological, spatial, and self-report data and to illustrate the types of insights that such integration can yield.

\subsubsection{Experience Sampling}

\begin{figure}[htbp]
\centering
\includegraphics[width=0.9\textwidth]{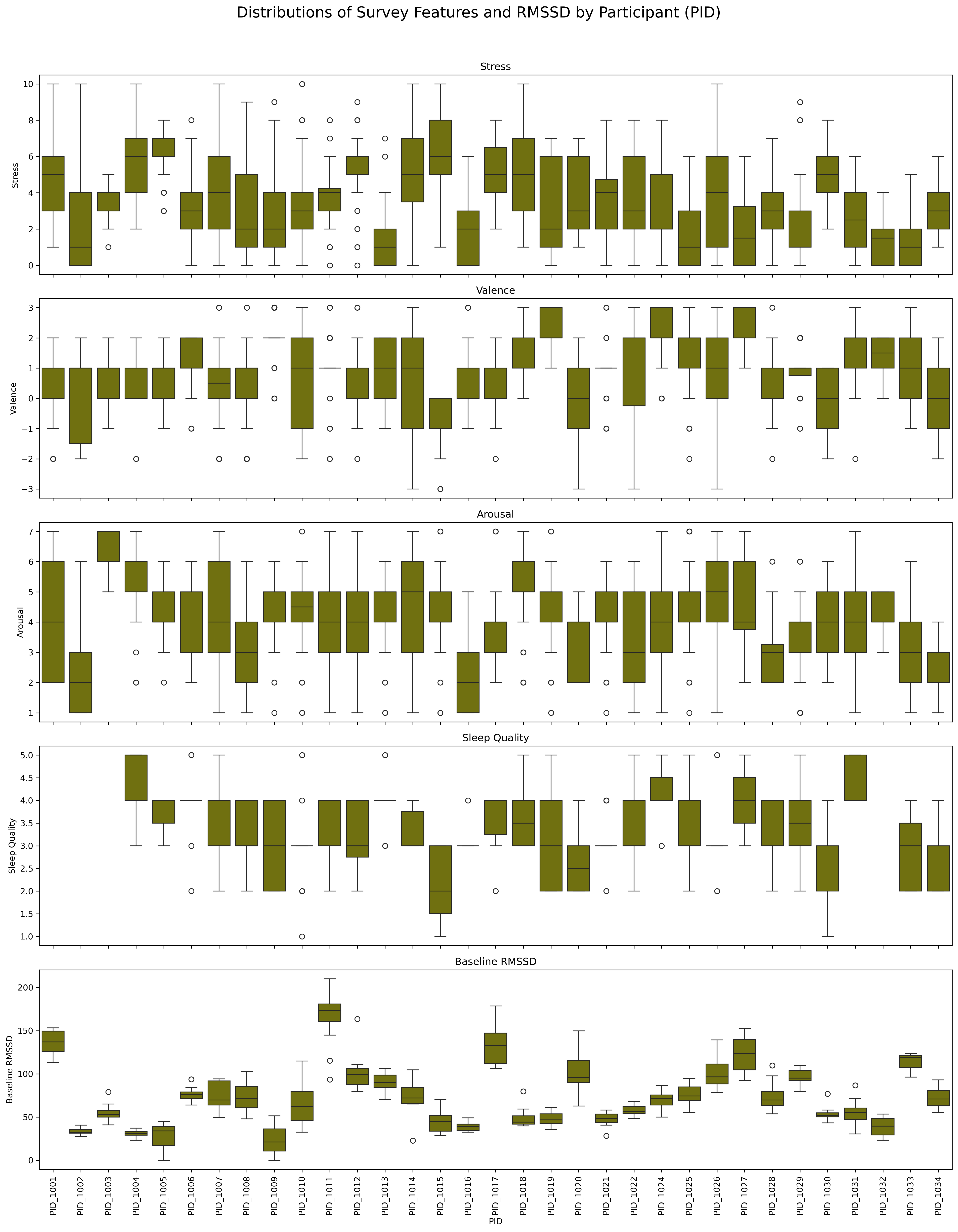}
\caption{Distributions of self-reported and physiological features across participants. Each subplot shows boxplots for one variable: stress, valence, arousal (sampled four times daily), sleep quality (sampled once daily), and baseline RMSSD computed from sleep periods. Each box represents within-participant variability across the 14-day study.}
\label{fig:boxplots}
\end{figure}

%-stress, affect, and these stuff. 
%-how much up and down. 
%-across participants how different they are. 
%-if walkign ahd anythign to do with stress/affect/sleep quality.
Figure~\ref{fig:boxplots} presents boxplots of five key variables, self-reported stress, valence, arousal, and sleep quality, as well as baseline RMSSD computed from sleep periods, plotted across all study participants (1001–1034). These visualizations summarize both within- and between-participant variability across the 14-day study period.

In the stress panel (top row of panels), ratings spanned the full 0–10 scale. Several participants, such as Participants 1007, 1014, and 1026 showed wide interquartile ranges (IQRs) and numerous outliers, indicating substantial variation in momentary stress across time. In contrast, participants 1013, 1032, and 1033 exhibited consistently low stress, with narrow IQRs and medians near zero.

The valence panel reveals a diverse range of emotional states. Participants 1019 and 1024 tended to report higher valence (median greater than 1), whereas Participant 1015 reported primarily negative valence. Some participants displayed symmetrical distributions around the median (e.g., 1034), while others showed strong skewness or greater dispersion.

The arousal panel illustrates considerable between-participant differences in perceived activation. High arousal variability is evident in participants 1018 and 1003, whereas participants 1017 and 1034 reported consistently low arousal across sampled events.

For sleep quality, most participants reported scores centered between 3 and 4 on the 1–5 scale. Participants 1004, 1024, and 1031 reported consistently high sleep quality, while others, including Participants 1009 and 1019, showed greater day-to-day variability.

The final panel shows baseline RMSSD distributions calculated from sleep periods. The interindividual variability was substantial: participants 1001, 1011, and  1017 had high baseline RMSSD values (median greater than 100 ms), while participants 1002, 1004, and 1016 exhibited lower values (median less than 50 ms). Within-participant RMSSD variability was generally low compared to the self-reported variables, although some individuals (e.g., participants 1010 and 1017) showed broader spreads.

Together, these plots illustrate the pronounced heterogeneity in both subjective experience and baseline parasympathetic activity, underscoring the importance of individualized baselines for interpreting physiological signals in daily life contexts.

\subsubsection{Walking Experience and Reported Infrastructure Problems}

To contextualize the daily variability in self-reported stress and well-being, we next examined their walking behaviors and experiences with the built environment. 

\begin{figure}[h]
\centering
\includegraphics[width=0.9\textwidth]{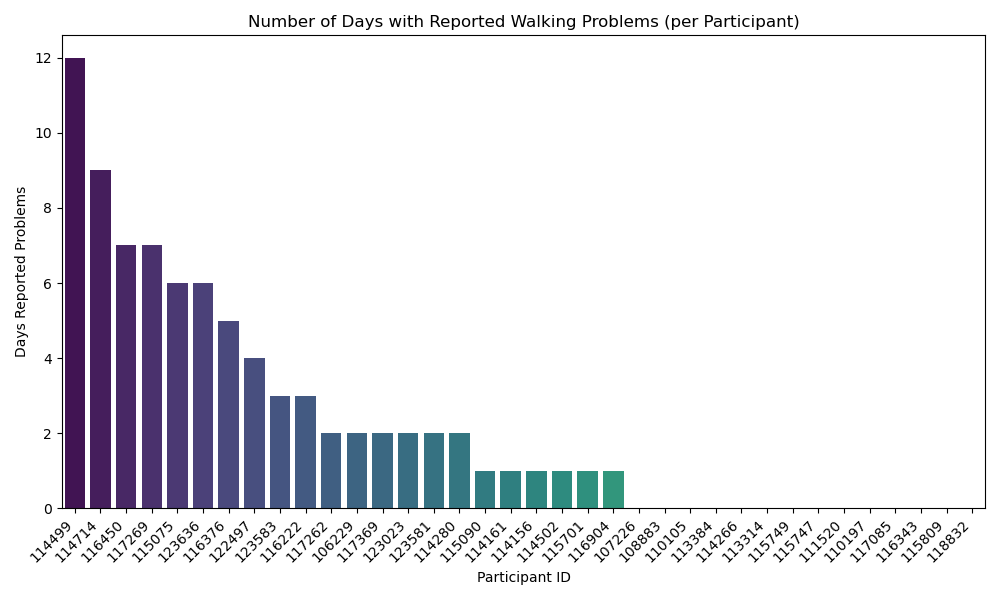}
\caption{Number of days each participant reported at least one pedestrian infrastructure problem. The figure shows variability in reporting frequency across individuals, with some participants reporting issues on up to 12 separate days and others reporting few or none.}
\label{fig:reported_problems_per_participant}
\end{figure}

Across the study period, most participants responded regularly to this prompt, generating a diverse set of reports describing obstacles and challenges encountered while walking. Figure~\ref{fig:reported_problems_per_participant} depicts the number of days each participant reported experiencing at least one walking-related problem. Overall, 61\% of participants reported encountering a problem at least once during the study period. A subset of participants reported issues on more than five days, with the participant showing the highest frequency logging infrastructure concerns on 12 separate days. In contrast, several participants reported problems only once or not at all. This variability likely reflects differences in walking routes, exposure to problematic infrastructure, as well as individual sensitivity and reporting tendencies.

\begin{figure}
\centering
\includegraphics[width=0.9\textwidth]{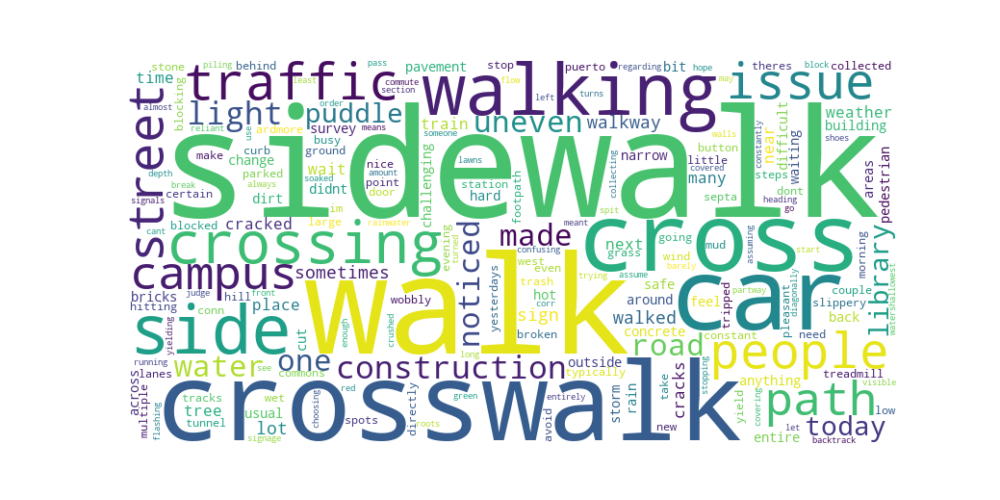}
\caption{Word cloud of participant-reported pedestrian infrastructure challenges. The visualization highlights frequently mentioned terms such as``sidewalk,'' ``crosswalk,'' ``walking,'' and ``traffic,'' as well as more specific issues like ``construction,'' ``uneven,'' ``puddle,'' and ``cracked.'' These terms reflect both common and nuanced concerns encountered across a variety of walking environments.}
\label{fig:word_cloud}
\end{figure}

We analyzed the open-ended responses using a text mining approach to identify common themes in participants’ walking experiences. To visualize the most frequently mentioned terms, we generated a word cloud based on the cleaned and aggregated text data. As shown in Figure~\ref{fig:word_cloud}, prominent words such as ``sidewalk,'' ``crosswalk,'' ``walking,'' and ``traffic'' appeared most frequently, indicating widespread concerns related to pedestrian infrastructure and road interactions. This visualization provides a high-level overview of the language participants used to describe their experiences, offering a qualitative snapshot of the recurring issues that may contribute to environmental stress during walking.

Beyond these dominant terms, participants’ responses revealed a variety of more specific issues that extended beyond general mentions of ``sidewalk'' or ``crosswalk.'' Responses ranged from references to surface conditions—such as ``cracked pavement,'' ``uneven bricks,'' and ``puddles,'' to broader concerns like ``construction zones,'' ``blocked paths,'' and ``poor lighting.'' Others described contextual factors including ``train stations,'' ``campus walkways,'' and ``narrow tunnels,'' highlighting the range of environments in which these walking challenges occurred. These detailed responses provide a more nuanced understanding of the pedestrian experience, capturing the diverse environmental and infrastructural factors that may influence physiological and emotional responses during urban walking.

\begin{figure}[h]
\centering
\includegraphics[width=0.9\textwidth]{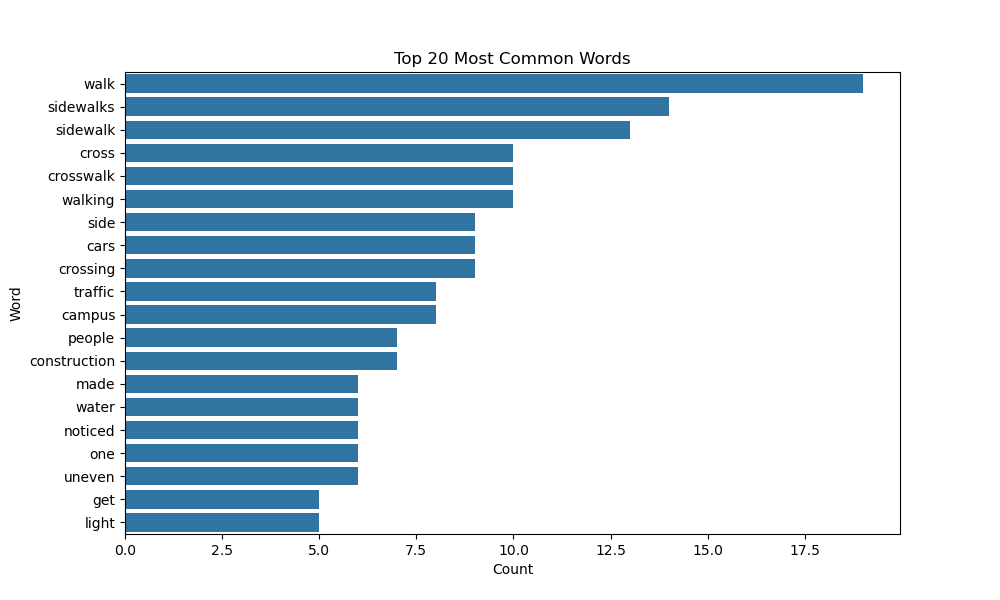}
\caption{Top 20 most common words mentioned in participants' descriptions of pedestrian infrastructure problems. While expected terms such as ``walk,'' ``sidewalk,'' and ``crosswalk'' dominate, the chart also highlights frequently reported issues including ``construction,'' ``traffic,'' ``uneven,'' and ``cars,'' reflecting both general and specific challenges encountered by participants.}
\label{fig:top_words}
\end{figure}

\begin{figure}[h]
\centering
\includegraphics[width=0.9\textwidth]{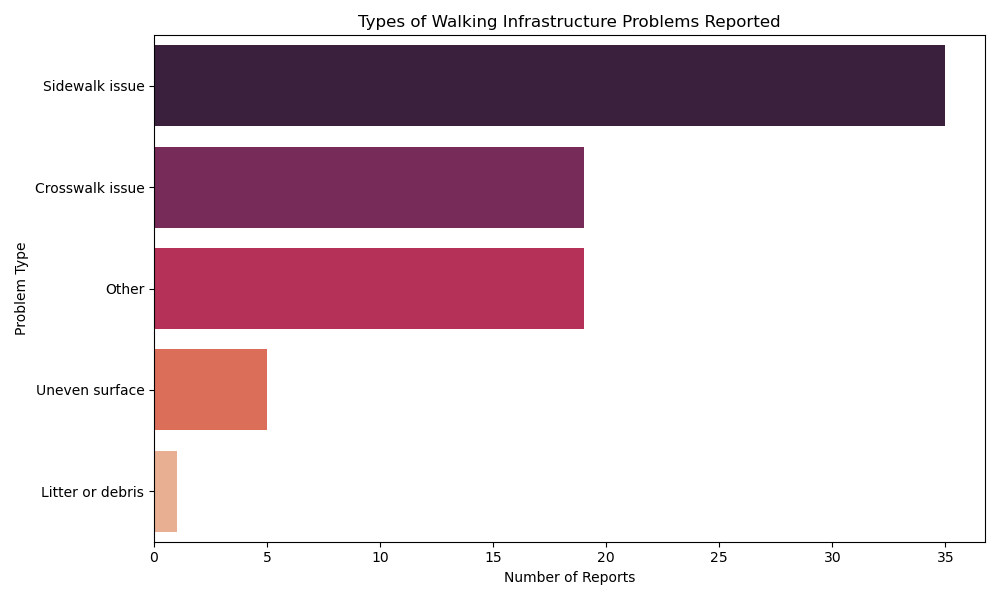}
\caption{Types of pedestrian infrastructure problems reported by participants. The most frequently reported issue was sidewalk-related, followed by crosswalk issues and open-text entries categorized as ``Other.'' Less common issues included uneven surfaces and litter or debris.}
\label{fig:problem_categories}
\end{figure}

Figure~\ref{fig:top_words} shows the 20 most frequently mentioned words across all participant responses. As expected, terms such as “walk,” “sidewalk,” and “crosswalk” appeared most often, underscoring the central focus on pedestrian experiences. Responses were categorized using keyword-based filtering (e.g., ``sidewalk,'' ``crosswalk,'' ``trash''), then manually reviewed and assigned to one of five predefined categories of ``Sidewalk Issues,'' ``Crosswalk Issues,'' ``Uneven Surfaces,'' ``Trash and Debris,'' and ``Others''. Figure ~\ref{fig:problem_categories} summarizes the types of pedestrian infrastructure problems reported using this method. 

The most frequently cited issue was sidewalk-related, accounting for more than one-third of all responses. These included broken or narrow sidewalks and fixed obstructions. One participant noted: ``Sometimes there are poles in the sidewalk that obstruct the ease of walking.''

Crosswalk issues were the second most common problem category. These frequently involved interactions with vehicle traffic and concerns about pedestrian safety. For example, one participant reported: ``Cars typically don't stop at a crosswalk I have to cross to get to the train station. I have to wait and hope there’s a break in traffic.''

Responses categorized as ``Other'' included open-text entries that did not fall under predefined categories but represented meaningful concerns. Notably, a recurring issue cited by multiple participants referenced a poorly maintained tunnel under train tracks near the university campus. For instance an individual mentioned ``The tunnel under the train tracks is difficult to go through because of the water on the ground constantly.''

Uneven walking surfaces and litter or debris were less frequently reported. While these categories occurred less often, they still reflect important environmental conditions that may affect walkability and comfort. These findings document the range of built environment challenges encountered by participants in their daily routines, highlighting pedestrian safety, accessibility, and infrastructure maintenance as recurring recurring concerns in urban walking contexts.

%\subsubsection{Physiological Data}

\subsubsection{Physiological Indicators of Stress During Walking}

Building on participants’ reports of pedestrian infrastructure challenges, we next conducted an observational analysis of autonomic activity during walking to examine how physiological arousal responses may correspond to environmental context. This analysis was not intended to represent all possible factors influencing stress, but rather to illustrate how these physiological measures can relate to the experiences reported by participants. As noted in the previous section, we analyzed two complementary measures of autonomic function: root mean square of successive differences (RMSSD) as a marker of parasympathetic regulation, and electrodermal activity (EDA) as an index of sympathetic arousal.

\begin{figure}[h]
\centering
\includegraphics[width=0.9\textwidth]{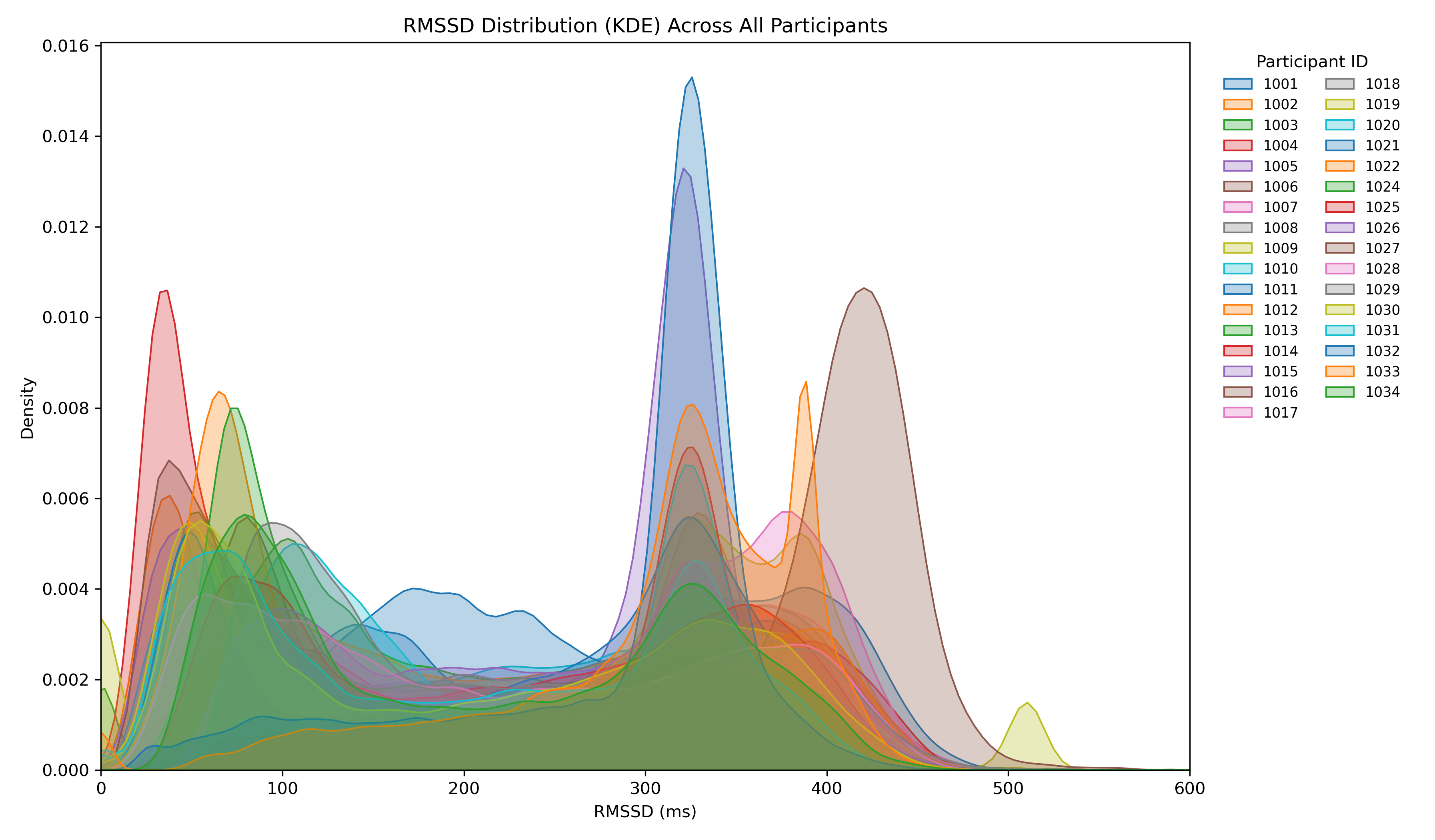}
\caption{Kernel density estimates of RMSSD for all participants, showing wide inter-individual variability in baseline autonomic activity.}
\label{fig:kde}
\end{figure}

Figure~\ref{fig:kde} shows kernel density estimates (KDEs) of RMSSD values for all 34 participants over their 14-day study period. Each curve represents a participant’s smoothed distribution of RMSSD values. The x-axis denotes RMSSD in milliseconds, and the y-axis represents relative density, normalized within each participant to illustrate the relative frequency of different RMSSD values.

RMSSD distributions varied widely across individuals. Some participants (e.g., 1001, 1021, and 1032) exhibited sharp unimodal peaks centered between 300–400 ms, indicating higher overall parasympathetic tone. Others (e.g., 1004, 1005, and 1020) had distributions concentrated below 100 ms. A few participants (e.g., 1011 and 1017) displayed broader or bimodal, suggesting greater within-person variation in RMSSD over time. Across all participants, RMSSD values ranged from approximately 25 to over 500 ms, reflecting wide interindividual differences in observed parasympathetic activity during walking.

\begin{figure}
\centering
\includegraphics[width=0.9\textwidth]{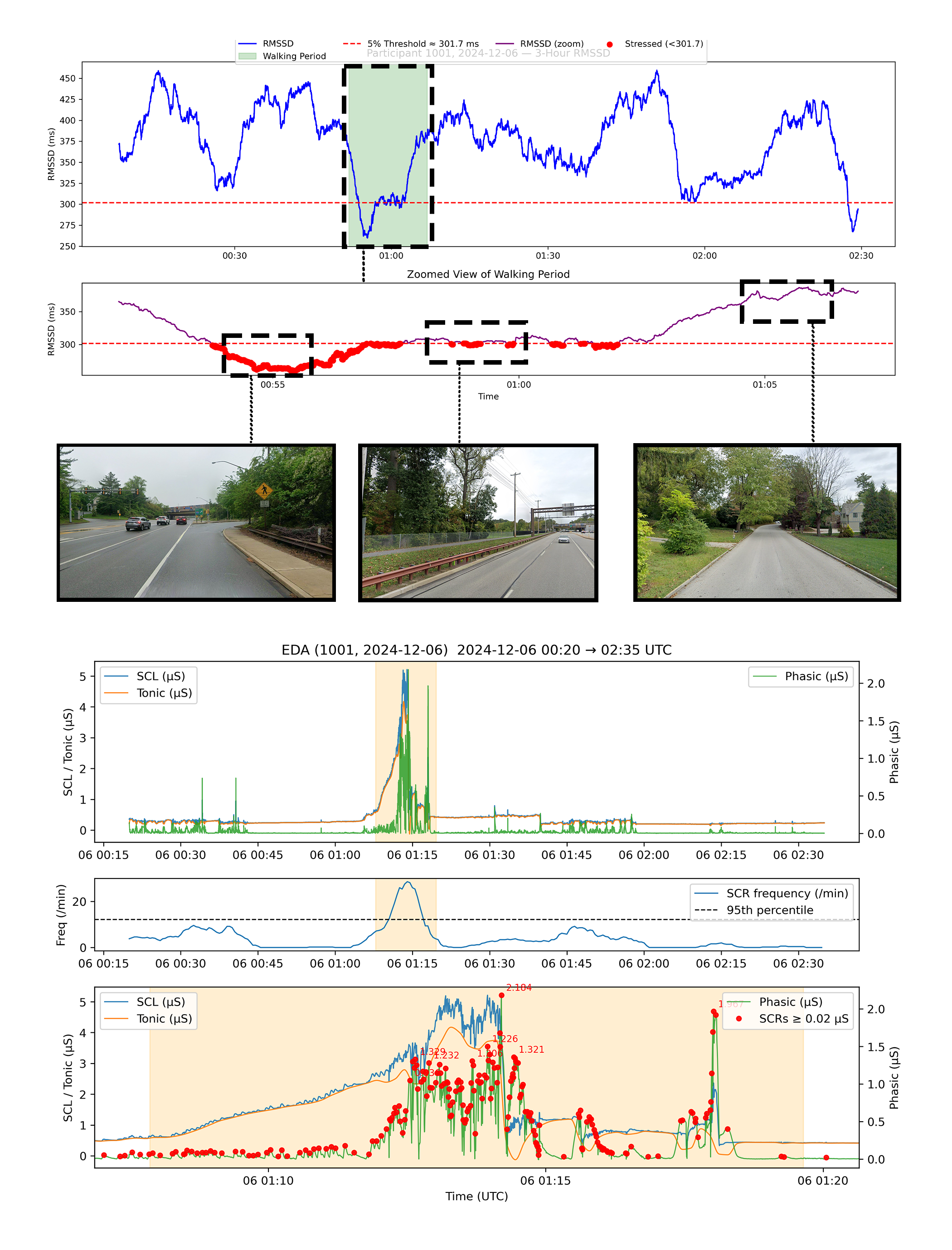}
\caption{RMSSD and EDA time series for Participant 1001 on December 6, 2024. The top panel shows the 3-hour RMSSD trace with the walking period (green) and the participant-specific stress threshold (red dashed line, 5th percentile $\approx$ 301.7 ms). The zoomed RMSSD segment highlights sub-threshold values in red. Lower panels show concurrent EDA signals, including tonic and phasic conductance, SCR frequency, and detected SCR events ($\geq$0.02 $\mu$S). Street-level images illustrate environmental context at three points along the route. Together, these multimodal data capture parasympathetic (RMSSD) and sympathetic (EDA) activity during real-world walking}
\label{fig:rmssd_eda_examples_a}
\end{figure}

\begin{figure}
\centering
\includegraphics[width=0.9\textwidth]{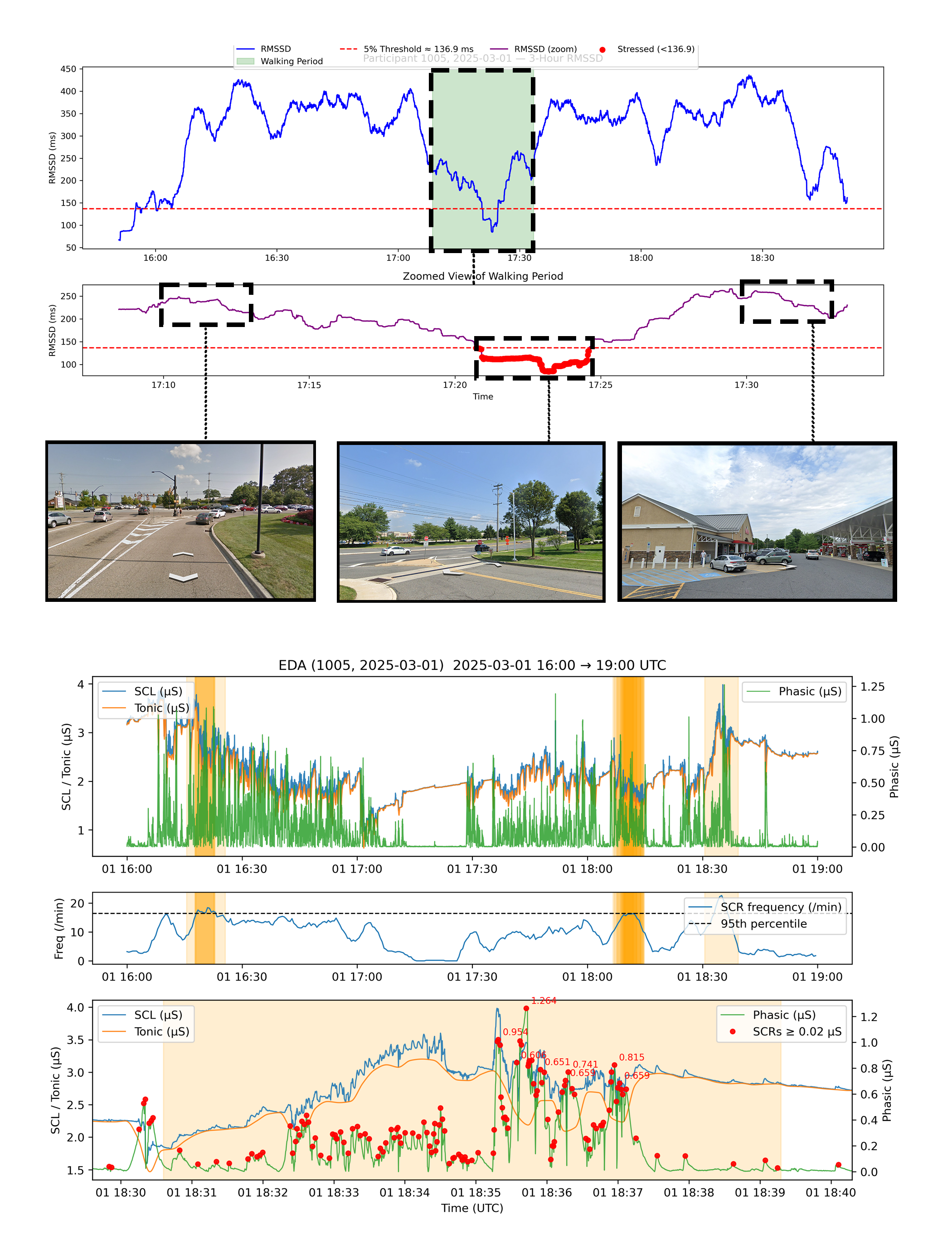}
\caption{RMSSD and EDA time series for Participant 1005 on March 1, 2025. The top panel shows the 3-hour RMSSD trace with the walking period (green) and the participant-specific stress threshold (red dashed line, 5th percentile $\approx$ 136.9 ms). The zoomed RMSSD segment highlights sub-threshold values in red. Lower panels show concurrent EDA signals, including tonic and phasic conductance, SCR frequency, and detected SCR events ($\geq$0.02 $\mu$S). Street-level images illustrate environmental context at three points along the route. Together, these multimodal data capture parasympathetic (RMSSD) and sympathetic (EDA) activity during real-world walking.}
\label{fig:rmssd_eda_examples_b}
\end{figure}

\begin{figure}
\centering
\includegraphics[width=0.9\textwidth]{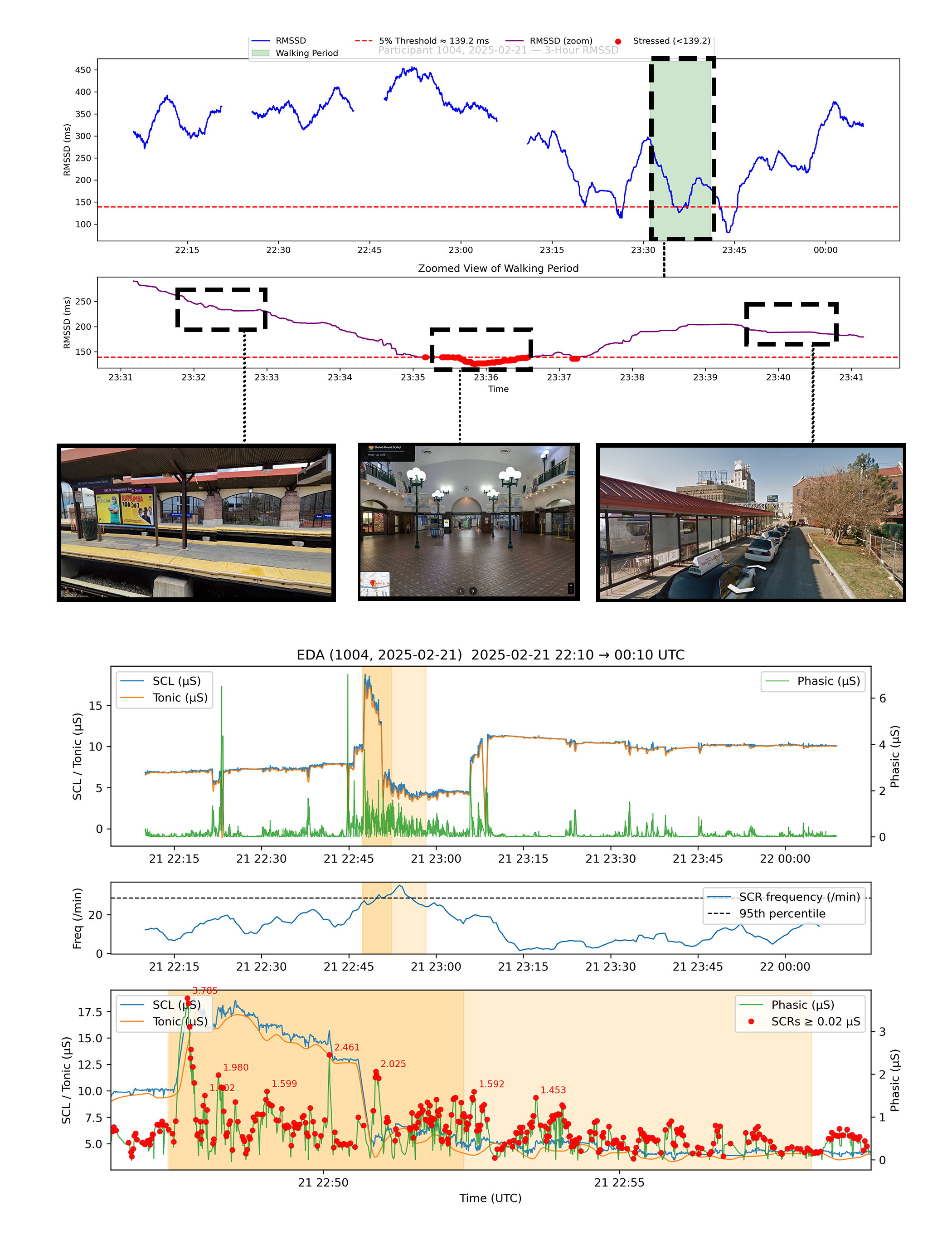}
\caption{RMSSD and EDA time series for Participant 1004 on February 21, 2025. The top panel shows the 3-hour RMSSD trace with the walking period (green) and the participant-specific stress threshold (red dashed line, 5th percentile $\approx$ 139.2 ms). The zoomed RMSSD segment highlights sub-threshold values in red. Lower panels show concurrent EDA signals, including tonic and phasic conductance, SCR frequency, and detected SCR events ($\geq$0.02 $\mu$S). Street-level images illustrate environmental context at three points along the route. Together, these multimodal data capture parasympathetic (RMSSD) and sympathetic (EDA) activity during real-world walking.}
\label{fig:rmssd_eda_examples_c}
\end{figure}

\begin{figure}
\centering
\includegraphics[width=0.9\textwidth]{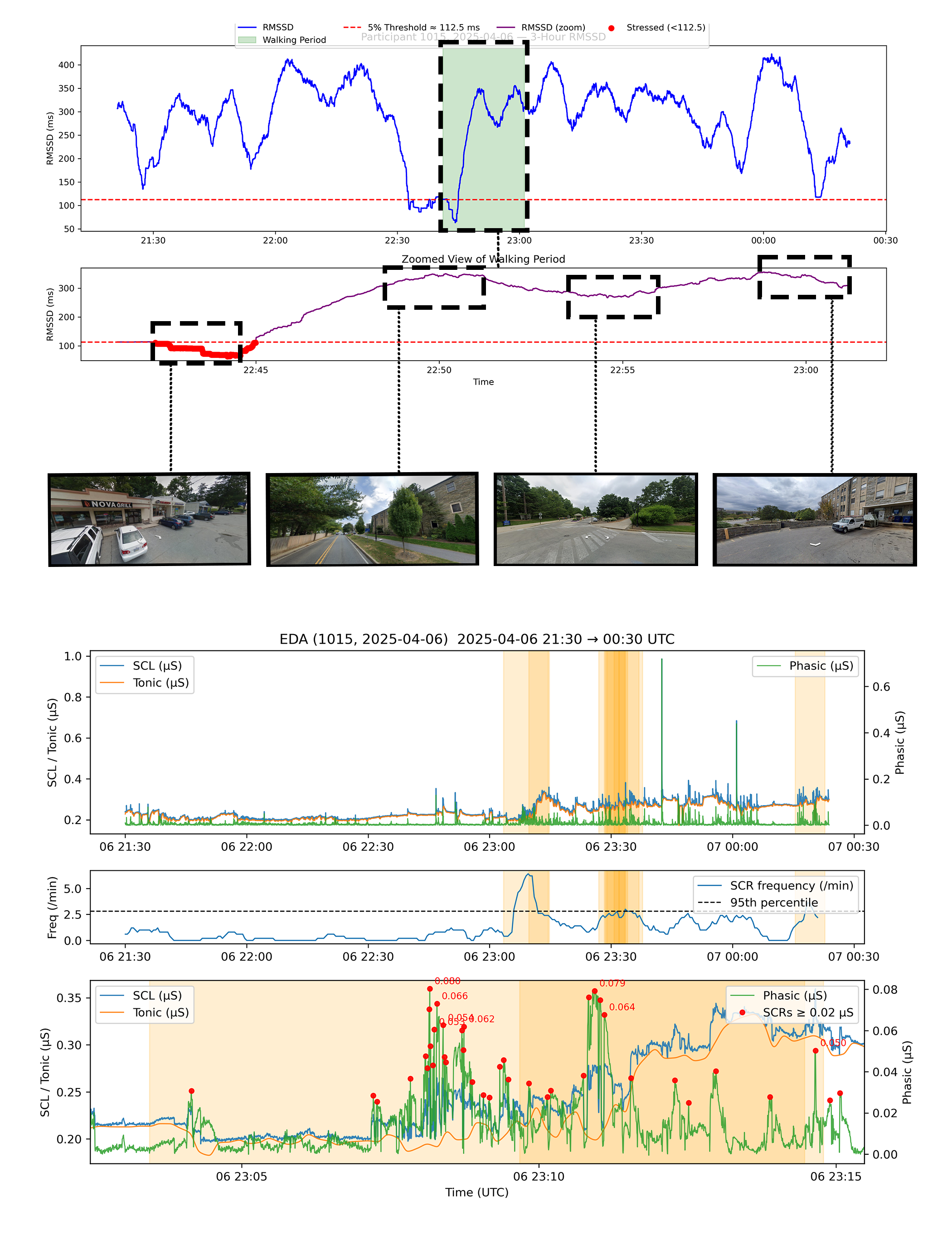}
\caption{RMSSD and EDA time series for Participant 1015 on April 6, 2025. The top panel shows the 3-hour RMSSD trace with the walking period (green) and the participant-specific stress threshold (red dashed line, 5th percentile $\approx$ 113 ms). The zoomed RMSSD segment highlights sub-threshold values in red. Lower panels show concurrent EDA signals, including tonic and phasic conductance, SCR frequency, and detected SCR events ($\geq$0.02 $\mu$S). Street-level images illustrate environmental context at three points along the route. Together, these multimodal data capture parasympathetic (RMSSD) and sympathetic (EDA) activity during real-world walking.}
\label{fig:rmssd_eda_examples_d}
\end{figure}

\begin{figure}
\centering
\includegraphics[width=0.9\textwidth]{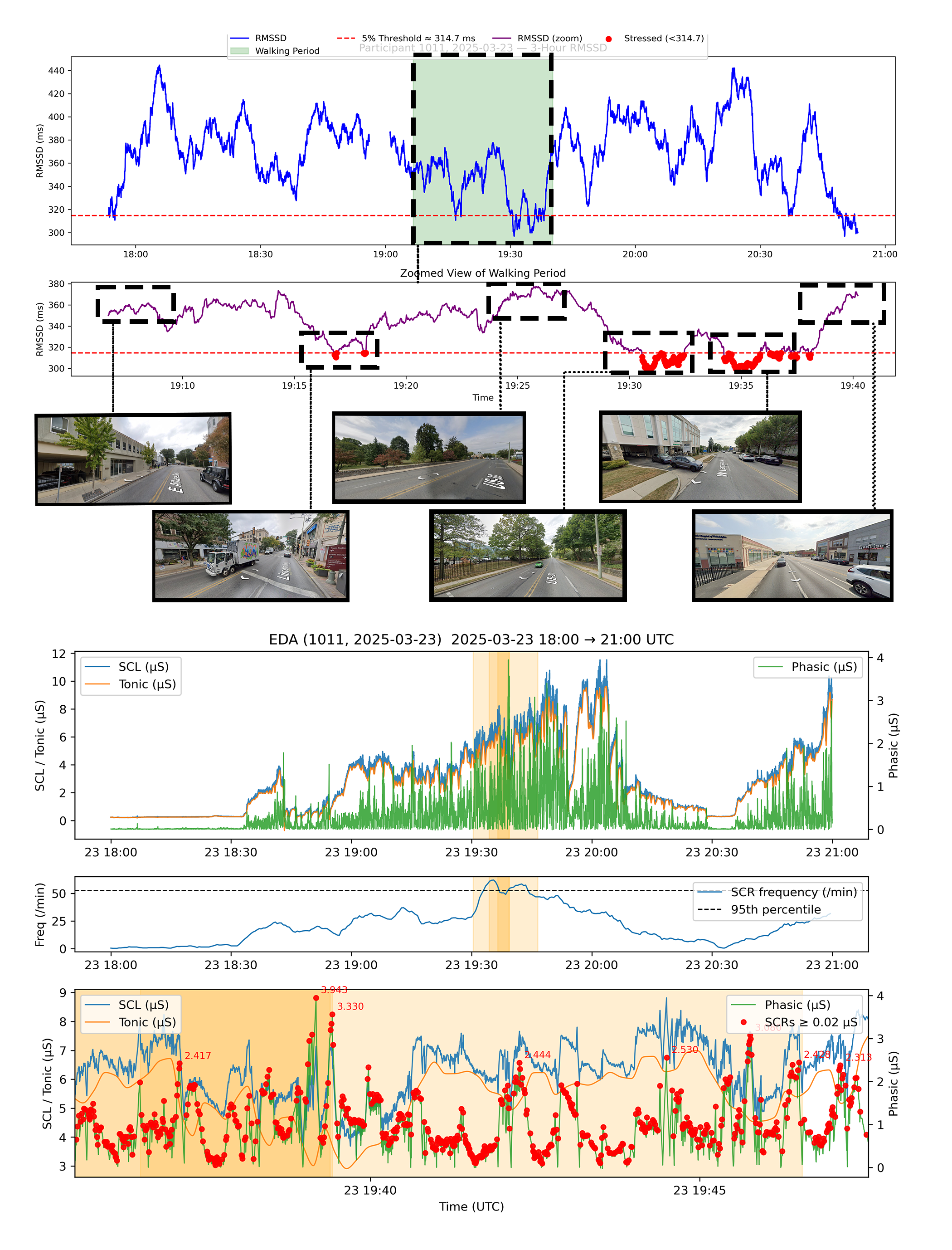}
\caption{RMSSD and EDA time series for Participant 1011 on March 23, 2025. The top panel shows the 3-hour RMSSD trace with the walking period (green) and the participant-specific stress threshold (red dashed line, 5th percentile $\approx$ 315 ms). The zoomed RMSSD segment highlights sub-threshold values in red. Lower panels show concurrent EDA signals, including tonic and phasic conductance, SCR frequency, and detected SCR events ($\geq$0.02 $\mu$S). Street-level images illustrate environmental context at three points along the route. Together, these multimodal data capture parasympathetic (RMSSD) and sympathetic (EDA) activity during real-world walking.}
\label{fig:rmssd_eda_examples_e}
\end{figure}

\begin{figure}
\centering
\includegraphics[width=0.9\textwidth]{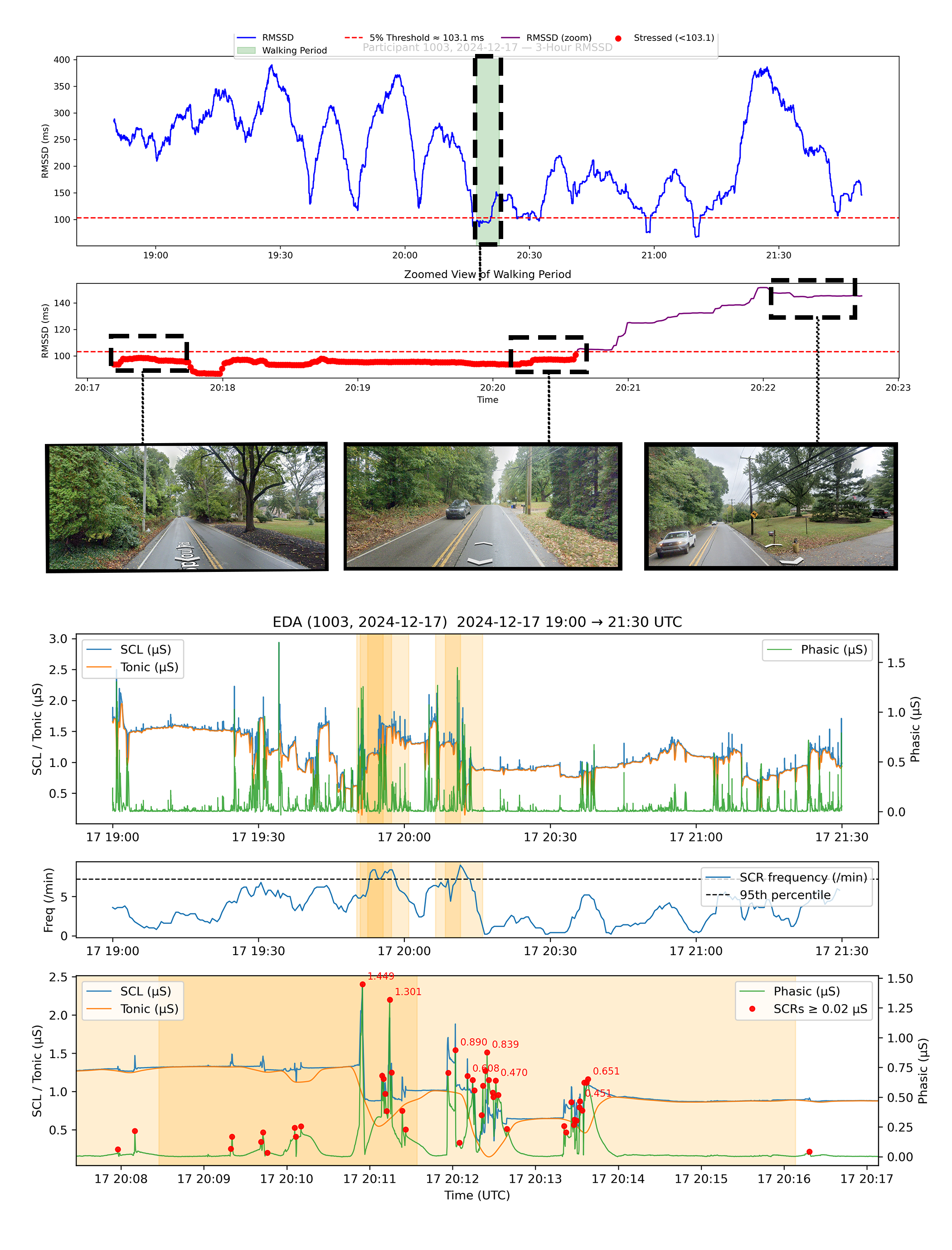}
\caption{RMSSD and EDA time series for Participant 1003 on December 17, 2024. The top panel shows the 3-hour RMSSD trace with the walking period (green) and the participant-specific stress threshold (red dashed line, 5th percentile $\approx$ 103 ms). The zoomed RMSSD segment highlights sub-threshold values in red. Lower panels show concurrent EDA signals, including tonic and phasic conductance, SCR frequency, and detected SCR events ($\geq$0.02 $\mu$S). Street-level images illustrate environmental context at three points along the route. Together, these multimodal data capture parasympathetic (RMSSD) and sympathetic (EDA) activity during real-world walking.}
\label{fig:rmssd_eda_examples_g}
\end{figure}

To explore how physiological responses during walking varied across environmental contexts, we analyzed RMSSD and EDA features in relation to participants' GPS location. To illustrate the types of data patterns observed, six representative participants were selected to demonstrate distinct examples of autonomic variation across different walking environments (Figures~\ref{fig:rmssd_eda_examples_a}–\ref{fig:rmssd_eda_examples_g}). As this article is focused on our framework, these cases are presented as descriptive observations intended to show the feasibility of multimodal integration rather than to infer causal associations.

Each composite visualization integrates multiple data streams to provide a detailed, time-synchronized view of physiological and contextual dynamics during walking. The top panel shows a 3-hour RMSSD time series in which walking periods, identified via GPS-derived speed (0.5–2.0 m/s), are highlighted in green. Below it, a zoomed view focuses on a single walking interval, with red segments indicating RMSSD values below the 5th percentile of that participant’s daily RMSSD distribution. This view emphasizes short-term decreases in RMSSD that may reflect transient changes in autonomic state during walking. 

The middle row presents three corresponding street-level images extracted from Google Street View that approximate the participants' location along the route. These contextual snapshots illustrate changes in the built environment, such as road width, the presence or absence of pedestrian buffers, vegetation, and proximity to intersections, that may coincide with fluctuations in physiological activity. 

The bottom panels depict electrodermal activity features over the same period. The first EDA panel shows tonic and phasic skin conductance components, providing a continuous record of slow- and fast-varying activity. The second panel displays the frequency of discrete skin conductance responses (SCRs) per minute, and the third panel overlays identified SCR events ($\geq$ 0.02 $\mu$S) on the phasic signal to indicate transient increases in skin conductance. These multimodal visualizations depict temporal co-variation between parasympathetic and sympathetic indicators during naturalistic walking, highlighting how the proposed framework can capture fine-grained physiological and contextual dynamics in everyday settings.

Participant 1001 (Figure~\ref{fig:rmssd_eda_examples_a}) exhibited a distinct drop in RMSSD below the 5th percentile threshold (301.7~ms) during a walking episode along a high-traffic arterial road characterized by minimal pedestrian buffering and low walkability (Walk Score = 39/100). The EDA trace showed a delayed increase in sympathetic indicators, with tonic conductance and phasic SCR activity rising several minutes after the RMSSD decrease. As the participant moved into a moderately walkable segment (Walk Score = 47/100) featuring partial separation from vehicular traffic, RMSSD values gradually increased. The highest RMSSD values and lowest SCR activity were observed in the final segment of the walk (Walk Score = 52/100), which occurred within a quiet residential street with greater green space and minimal traffic exposure. The sequential increase in walkability scores along this route coincided with gradual recovery in RMSSD, which may indicate that features of more walkable or comfortable environments were associated with more stable autonomic activity for this participant.

Participant 1005 (Figure~\ref{fig:rmssd_eda_examples_b}) exhibited a pronounced reduction in RMSSD while walking through a commercial corridor and adjacent parking lot, reaching a minimum of approximately 120~ms around 17:20~UTC. This route, located in a moderately walkable area (Walk Score = 55/100), included limited pedestrian infrastructure and exposure to nearby vehicular traffic. During this interval, the EDA signal was absent, likely due to temporary loss of sensor contact or environmental interference, preventing concurrent observation of sympathetic activity. When the EDA signal resumed later in the recording (around 17:30~UTC), a cluster of phasic SCR events with amplitudes exceeding 1~$\mu$S was detected near 18:30~UTC, accompanied by increased SCR frequency. Although this EDA activity occurred nearly an hour after the observed RMSSD reduction, the temporal separation highlights the complexity of interpreting multimodal physiological data in real-world settings and the influence of both environmental variation and sensor reliability on observed patterns.

%The middle panel provides a zoomed view of the walking episode, highlighting sub-threshold RMSSD values in red. RMSSD declines steadily starting just before 00:55 and begins to recover after 01:05, with a sustained sub-threshold period of approximately 10 minutes.

%Street-level images in the bottom row correspond to three points along the walking route. The lowest RMSSD values occurred while the participant walked along a sidewalk directly adjacent to traffic, with no physical separation such as fencing or greenery. RMSSD gradually increased as the participant transitioned to a sidewalk with more visual and physical separation, and peaked near the end of the walk in a quiet suburban neighborhood. This pattern aligns reductions in parasympathetic activity with specific changes in built environment features.

.
%A second example, shown in Figure~\ref{fig:rmssd2}, is from Participant 1005 on March 1, 2025. RMSSD begins to decline around 17:14 as the participant walks along a multi-lane commercial road with no physical barrier between the sidewalk and traffic (left image). The lowest RMSSD values are observed between 17:20 and 17:25, while the participant crosses an open intersection and navigates the edge of a large parking lot (middle image). This area lacks designated pedestrian pathways. RMSSD gradually increases after 17:25 as the participant approaches the entrance to a commercial plaza, where walkways and curb ramps are more clearly defined (right image). As with the previous example, the physiological signal aligns with shifts in environmental structure and exposure to traffic.

Participant 1004 (Figure~\ref{fig:rmssd_eda_examples_c}) exhibited a clear temporal offset between sympathetic and parasympathetic indicators while traveling through a transit hub and adjacent underpass. The location was situated in a highly walkable area (Walk Score = 94/100), characterized by dense land use, multimodal connectivity, and proximity to public transit. The EDA signal showed a pronounced increase beginning around 22:50~UTC, coinciding with the period when the participant was on or exiting the train. During this interval, tonic skin conductance level rose above 10~$\mu$S, accompanied by several high-amplitude SCR bursts (up to 2–3~$\mu$S) and elevated SCR frequency. These observations suggest a period of heightened sympathetic activity that coincided with the transition through the transit environment.

In contrast, the RMSSD trace revealed a subsequent decline in parasympathetic activity beginning around 23:30~UTC, falling below the participant-specific 5th percentile threshold (approximately 139~ms) between 23:33 and 23:39. This reduction occurred while the participant was walking through the train station and adjacent underpass—locations characterized by enclosed corridors and concentrated pedestrian flow despite high functional walkability.

The temporal separation between EDA and RMSSD changes indicates asynchronous dynamics across autonomic branches for this participant: sympathetic activity increased earlier, whereas parasympathetic suppression appeared later during the station traversal. When considered alongside the corresponding environmental imagery, these observations illustrate how localized design features, such as enclosure, crowd density, or limited visual openness, can coincide with distinct temporal patterns in physiological measures, even within otherwise walkable urban areas.

Participant 1015 (Figure~\ref{fig:rmssd_eda_examples_d}) exhibited a brief episode of autonomic variation during an evening walk through a mixed-use suburban area. RMSSD values declined rapidly between 22:40 and 22:50~UTC, falling below the participant-specific 5th percentile threshold (approximately 113~ms) while walking along a narrow, traffic-adjacent street with parked vehicles and moderate walkability (Walk Score = 44/100). This short-lived reduction in RMSSD represents a momentary decrease in parasympathetic activity that coincided with exposure to a vehicle-dense environment and limited pedestrian separation.

The EDA signal displayed a more gradual and delayed change. Beginning near 23:00~UTC, tonic skin conductance level (SCL) increased slightly, accompanied by a cluster of small phasic SCRs (0.05–0.08~$\mu$S) and a modest rise in SCR frequency. The temporal offset between RMSSD and EDA fluctuations suggests asynchronous dynamics across autonomic branches for this participant, where a reduction in parasympathetic activity was followed by a mild sympathetic increase several minutes later.

The contextual imagery shows a transition from a busy roadside segment (Walk Score = 44/100) and a secondary street with similarly low walkability (Walk Score = 41/100) into pedestrian-oriented university grounds, where no formal Walk Score was available but infrastructure was designed primarily for foot traffic. This environmental change coincided with a gradual increase in RMSSD and stabilization of EDA activity after 22:50~UTC, illustrating how shifts in walking context corresponded with short-term variations in autonomic indicators during everyday mobility.

Participant 1011 (Figure~\ref{fig:rmssd_eda_examples_e}) exhibited distinct but temporally offset patterns in parasympathetic and sympathetic indicators while walking along a busy multi-lane urban street. RMSSD values began declining around 19:25~UTC and remained below the participant-specific threshold (approximately 315~ms) for nearly ten minutes, representing a sustained reduction in RMSSD during this period. This reduction occurred primarily within a moderately less walkable segment of the route (Walk Scores = 77–83/100), where the participant traversed a section characterized by narrow sidewalks, limited pedestrian buffering, and close proximity to vehicular traffic.

The EDA signal peaked later, around 19:40~UTC, after RMSSD values had begun to increase. During this interval, tonic skin conductance level exceeded 8~$\mu$S, accompanied by frequent, high-amplitude phasic responses (up to 3–3.5~$\mu$S) and elevated SCR frequency above the 95th percentile threshold. The temporal offset between RMSSD and EDA activity suggests asynchronous fluctuations in parasympathetic and sympathetic measures for this participant, with sympathetic activity rising several minutes after RMSSD recovery.

The environmental imagery and walkability data indicate that both the preceding and subsequent route segments, located in denser commercial areas with higher walkability (Walk Scores = 92–93/100), corresponded to relatively more stable RMSSD and EDA profiles. For this participant, these observations illustrate that short-term variations in autonomic activity may coincide with local design features—such as sidewalk width, traffic exposure, and sensory intensity, even within urban areas that are otherwise rated as highly walkable.

Participant 1003 (Figure~\ref{fig:rmssd_eda_examples_g}) exhibited a clear temporal offset between sympathetic and parasympathetic indicators during an evening walk along a narrow, tree-lined residential road. The EDA signal showed a pronounced increase in activity between 20:08 and 20:16~UTC, characterized by elevated phasic responses and several SCR events exceeding 1~$\mu$S in amplitude. During this interval, SCR frequency rose above the 95th percentile threshold, suggesting a period of sustained sympathetic activation. Tonic skin conductance also increased modestly, consistent with elevated overall physiological engagement.

Shortly after this EDA increase, RMSSD began to decline around 20:17~UTC, falling below the participant-specific 5th percentile threshold (approximately 103~ms) for roughly two minutes. This sequence, an initial increase in sympathetic indicators followed by a brief reduction in RMSSD, suggests temporally asynchronous changes in autonomic measures for this participant.

The spatial context shows that this episode occurred within a low-walkability environment (Walk Scores = 8–11/100) with minimal pedestrian infrastructure, absence of sidewalks, and close proximity to vehicular lanes. These localized environmental characteristics coincided with the observed fluctuations in EDA and RMSSD. Both measures returned to baseline shortly thereafter, indicating a short-lived deviation in autonomic activity during this portion of the walk.

%-HRV
%-when they ahd stress when they did not. RMSSD 
%-map to location. 
%-WHich locations they had stress
%-You can go back to those location on google map 
%-you can also look at APIs
%

%-for EDA same thing

\section{Discussion}

This study introduced a human-centered, multimodal framework for examining pedestrian well-being and interactions with urban infrastructure by integrating physiological signals, self-reported experiences via the Experience Sampling Method, and geospatial data from GPS. By deploying this framework in a naturalistic urban setting, we illustrate its feasibility and its potential to capture dynamic, context-sensitive patterns of pedestrian experience. The integration of multiple data streams allows for moving beyond static or aggregate assessments of urban environments, providing fine-grained, time-resolved observations of how individuals experience their surroundings. In the following, we reflect on key observations, consider their implications for research and practice, and outline directions for future work.

%Implementing the multimodal framework on a city-wide scale offers a transformative, cost-effective alternative to traditional urban planning audits. While research-grade wearables like the Empatica EmbracePlus are used for high-fidelity validation, the system is designed to be hardware-agnostic, allowing for potential large-scale integration with common consumer smartwatches. By leveraging the ubiquitous presence of smartphones and wearable tech, city officials can move beyond static walkability indices to capture real-time, high-resolution human sensors. This "participatory sensing" approach allows planners to identify fine-scale infrastructure issues—such as specific unlit tunnels, narrow sidewalk segments, or hazardous crosswalks—that are often invisible in aggregate data but trigger measurable physiological stress in residents. Ultimately, this granular data enables more inclusive, human-centered urban design, prioritizing repairs and safety interventions where they most directly improve the collective well-being of the population.

Implementing the proposed multimodal framework at a city-wide scale could provide a practical and cost-efficient complement to traditional urban planning audits. Although research-grade wearables such as the Empatica EmbracePlus were used in this study to ensure high-fidelity data collection, the framework is designed to be hardware-agnostic and could be adapted for broader deployment using commonly available consumer smartwatches. By leveraging the widespread adoption of smartphones and wearable technologies, municipal planners could supplement existing walkability indices with dynamic, fine-grained indicators derived from real-time physiological and experiential data. This participatory sensing approach may help identify localized infrastructure challenges, such as poorly lit underpasses, narrow sidewalks, or unsafe crosswalks, that are often overlooked in aggregate datasets yet may influence pedestrian comfort and safety. In this way, such data could support more inclusive and human-centered urban design practices, guiding maintenance and safety improvements toward locations where they are most likely to enhance everyday well-being.

The data revealed substantial heterogeneity in participants’ self-reported psychological states, including perceived stress, affective valence, arousal, and sleep quality, both across and within individuals. While such variability is expected in naturalistic settings, its structured capture through ESM provides a means of quantifying daily fluctuations in subjective experience \cite{mill2016emotional}. Similarly, physiological measures exhibited marked inter- and intra-individual variability. RMSSD, used here as an index of parasympathetic activity, varied widely across participants—from consistently elevated baselines to more dynamic, multimodal distributions. Electrodermal activity (EDA) complemented this pattern by showing both transient and sustained fluctuations in sympathetic activation that were sometimes, but not always, temporally aligned with RMSSD variations. Prior work has shown that EDA and HRV signals may not evolve synchronously because sympathetic influences reach peripheral effectors at different rates. In particular, under heightened arousal, sweat gland activation and cardiovascular adjustments can occur on distinct temporal scales, producing observable delays between EDA and RMSSD changes \cite{zhang2023dynamic, alberdi2016towards}.

The multimodal patterns observed in this study build upon prior research on affective dynamics and walking-related stress monitoring \cite{kuppens2010feelings}, suggesting that sympathetic (EDA) and parasympathetic (RMSSD) responses may differ in timing and magnitude during everyday walking \cite{kerous2020examination, ghiasi2020assessing}. In several participants, increases in EDA preceded reductions in RMSSD, suggesting a possible sequence of heightened sympathetic activation followed by reduced parasympathetic tone. In other cases, both measures varied in parallel, reflecting concurrent changes across autonomic branches. These temporally asynchronous or coordinated patterns highlight the importance of analyzing both systems jointly when characterizing physiological responses to environmental exposure \cite{visnovcova2013heart, kim2020electrogastrogram}. Overall, the results illustrate that autonomic changes in naturalistic contexts can occur on distinct timescales and may represent complex, context-dependent physiological adjustments rather than uniform or simultaneous responses.

Environmental context appeared to coincide with participants’ physiological responses in observable ways. Sub-threshold RMSSD events and heightened EDA activity were often recorded near urban features such as high-traffic corridors, unbuffered sidewalks, open intersections, and enclosed or poorly lit spaces. For example, participants walking through transit hubs or commercial streets exhibited short-term RMSSD declines and increased SCR frequencies, patterns consistent with elevated environmental stimulation. In contrast, gradual RMSSD recovery and lower EDA activity were more commonly observed when participants entered quieter or more visually enclosed residential areas. These descriptive patterns align with prior work in urban design and environmental psychology linking noise, vehicular proximity, and limited pedestrian separation to higher perceived stress and reduced sense of safety \cite{dumbaugh2005safe,vallejo2020perception}. Because the present study involved a small sample and non-experimental design, these associations should be interpreted as illustrative observations rather than generalizable or causal effects.

The observed physiological sequences also highlight both the challenges and the advantages of real-world multimodal data collection. Instances of signal dropout, such as transient EDA loss during walking, underscore the technical and environmental limitations of continuous sensing in naturalistic contexts. Despite these constraints, integrating RMSSD, EDA, imagery, and location data provided an ecologically valid means of contextualizing autonomic variation during walking. These case analyses illustrate that meaningful physiological patterns can be captured and interpreted even under complex, uncontrolled environmental conditions, though further methodological refinement is needed to improve data completeness and synchronization.

Findings from this exploratory deployment also suggest that the EPA Walkability Index, while a useful infrastructure-based proxy for pedestrian accessibility, may not fully capture the experiential dimensions of walkability. In several cases, participants exhibited elevated physiological arousal even in areas with high walkability ratings (above 90/100). While other unobserved factors could have contributed to these responses, this pattern may indicate that formal walkability indices overlook psychosocial and perceptual aspects of pedestrian experience \cite{venerandi2024walkability, knapskog2019exploring}. Perceptions of safety, environmental cleanliness, and sensory comfort likely play important roles in shaping pedestrian well-being beyond what infrastructure metrics alone can explain \cite{forsyth2015walkable, beil2013influence, tavakoli2025psycho}. Conversely, areas with lower walkability scores were more frequently associated with sub-threshold RMSSD events and elevated sympathetic activity, consistent with prior studies linking infrastructural deficits such as narrow sidewalks and high traffic exposure to physiological strain \cite{ewing2009measuring}. Future research could extend these findings by combining subjective perceptions and fine-grained environmental indicators, such as noise, air quality, or crowd density, to build more comprehensive and human-centered models of walkability.

ESM further enhanced the framework by providing subjective context for interpreting physiological variation. Open-text responses revealed recurring concerns such as sidewalk obstructions, crosswalk safety, and poorly maintained infrastructure, echoing environmental features that coincided with autonomic changes. Integrating these qualitative reports with physiological and spatial data offers a more holistic view of pedestrian well-being, linking lived experience with objective measurement. Such multimodal integration may help identify recurring environmental stressors that physiological signals alone cannot fully capture.

Collectively, these results suggest that multimodal, human-centered sensing approaches hold promise for advancing the study of environmental influences on well-being. The framework presented here enables the concurrent examination of immediate physiological responses and longer-term subjective experiences, offering a foundation for analyzing how environmental transitions correspond with changes in autonomic balance. The individualized baseline calibration and joint monitoring of parasympathetic and sympathetic activity provide methodological groundwork for future adaptive and personalized urban stress models. Future studies incorporating additional environmental sensing modalities could yield more precise characterizations of how moment-to-moment urban exposures relate to human physiology and affective states in situ.

Finally, the proposed multimodal framework has potential applications across multiple disciplines concerned with human experience in urban contexts. Urban planners and designers may use such approaches to identify location-specific stressors and inform interventions that enhance pedestrian comfort and inclusivity. Public health researchers could apply this framework to monitor physiological and behavioral markers of well-being across populations, supporting finer-grained assessments of environmental impacts on health. Behavioral scientists might leverage multimodal data to investigate real-world emotional and cognitive states in different spatial settings. Likewise, transportation engineers and policy-makers could employ these tools to evaluate infrastructure performance through human-centered indicators—such as emotional strain, perceived safety, or cognitive demand—alongside traditional mobility metrics. The adaptability of the framework across sensing platforms and survey tools underscores its potential to support diverse research and applied objectives, while ongoing refinement and validation will be essential to ensure reliability and interpretive robustness.

\section{Limitations and Future Work}

While the proposed multimodal framework provides a promising foundation for capturing momentary psychological and physiological states in naturalistic urban environments, several limitations warrant careful consideration and provide directions for future research.

The relatively small and homogeneous sample size may limit the generalizability of the findings. Although the idiographic design allowed for detailed within-person analysis and temporal resolution, the sample, drawn primarily from a university population, does not represent the diversity of age, socioeconomic status, or mobility behavior present in the general urban population. Future studies should recruit larger and more heterogeneous samples to capture variability across demographic, geographic, and cultural contexts, enabling stronger population-level inferences and cross-city comparisons.

Although this study examined both parasympathetic (RMSSD) and sympathetic (EDA) indicators, interpretation of autonomic activity remains constrained by the temporal and contextual complexity of these signals. RMSSD and EDA are influenced by multiple factors beyond psychological stress, including physical exertion, temperature, hydration, and circadian rhythms. Moreover, the two systems do not always respond synchronously, delays between sympathetic activation and parasympathetic withdrawal complicate direct interpretation of causality. Incorporating additional physiological channels such as respiration rate, skin temperature dynamics, or frequency-domain HRV metrics (e.g., LF/HF ratio) could improve temporal resolution and support a more comprehensive understanding of autonomic regulation in real-world contexts.

The percentile-based RMSSD thresholding and event-level EDA classification approaches, while effective for individualized detection of autonomic arousal, remain sensitive to confounding effects of movement and environmental variation. Future implementations could integrate accelerometry, posture recognition, or contextual variables such as ambient temperature and noise levels to distinguish between physiological arousal driven by environmental stressors and that arising from physical exertion or thermoregulation. Such enhancements would increase both the interpretive clarity and ecological validity of multimodal analyses.

While this study incorporated the EPA Walkability Index as an environmental descriptor, this metric primarily captures infrastructural and accessibility features rather than perceptual or affective qualities of the walking experience. The environmental analyses presented here were exploratory and intended as proof of concept. Future work could expand environmental characterization by integrating higher-resolution spatial data (e.g., air quality, sound levels, light intensity, or crowd density) alongside computer vision analysis of street-level imagery to capture visual and sensory dimensions of walkability. Combining these measures with physiological and self-report data could facilitate a more holistic understanding of how built environment features shape subjective well-being.

The Experience Sampling Method, while valuable for minimizing recall bias and improving ecological validity, may introduce participant burden through frequent survey prompts. Although compliance in this study was generally acceptable, response fatigue and missing entries could have influenced data completeness. Adaptive or context-triggered sampling strategies may offer a more participant-friendly and efficient alternative in future deployments.

Finally, the study’s short observation window and focus on a single urban region limit the temporal and spatial generalizability of the findings. Extending data collection over longer durations, seasons, and multiple cities would enable examination of how daily rhythms, weather variability, and regional design typologies influence pedestrian experience.

Looking ahead, the next phase of research could focus on scaling the multimodal framework in three key directions. First, methodological refinement should include the development of open-source pipelines and automated data alignment tools to improve reproducibility and facilitate comparison across studies. Second, participatory expansion could involve deploying the framework in diverse urban settings, allowing cities to collect fine-grained insights into pedestrian comfort and safety that complement traditional planning audits. Third, predictive modeling can integrate multimodal data streams into machine learning systems to infer pedestrian well-being in real time. Such models could support responsive urban interventions, including adaptive lighting, dynamic routing, or context-aware alerts. Collectively, these efforts would advance the framework from a proof of concept to a scalable and ethically grounded tool for understanding and enhancing well-being within everyday urban mobility.

\section{Conclusion}

This paper presents a human-centered, multimodal framework for assessing pedestrian well-being in real-world urban contexts. By integrating physiological data from wearable sensors, momentary self-reports via the Experience Sampling Method, and location traces from GPS, the framework enables nuanced exploration of how pedestrians experience their environments. In contrast to prior studies that often rely on predefined walking routes or retrospective surveys, this framework captures spontaneous, in-the-wild behavior during participants’ actual daily routines. This allows for ecologically valid insights into how environmental and infrastructural conditions shape both subjective and physiological responses. Built using commercially available tools and deployable applications, the framework is scalable and adaptable for broader urban research. Through a case study combining RMSSD and EDA-derived physiological arousal, ESM responses, and geolocated infrastructure problem reports, we demonstrate the feasibility of synchronizing multimodal data streams to reveal meaningful patterns in pedestrian experience. This work provides a foundation for future research on urban design, and individualized models of well-being.

\section{Acknowledgment}

This material is based upon work supported by the National Science Foundation under Grant No. \#2347012. Any opinions, findings, and conclusions or recommendations expressed in this material are those of the author(s) and do not necessarily reflect the views of the National Science Foundation.

\bibliographystyle{elsarticle-num-names} 
\bibliography{cas-refs}

%% else use the following coding to input the bibitems directly in the
%% TeX file.

%% Refer following link for more details about bibliography and citations.
%% https://en.wikibooks.org/wiki/LaTeX/Bibliography_Management

\end{document}